\def\exp{\hbox{exp}}
\def\ln{\ell{{\hbox {n}}}}
\documentstyle[a4,12pt]{article}

\newcommand{\beq}{\begin{equation}}
\newcommand{\eeq}{\end{equation}}
\def\beqa#1{\beq \begin{array}{#1}}
\newcommand{\eeqa}{\end{array} \eeq}

\newcommand{\xv}{\vec{x}}

\newcommand{\de}{\partial}
\newcommand{\De}{\nabla}

\newcommand{\ri}{\right}
\newcommand{\lt}{\left}
\newcommand{\ov}{\over}
\newcommand{\lp}{l(l+1)}

\def\reff#1{(\ref{#1})}				
\def\g^a#1#2#3{\Gamma^{{#1}{#2}}_{#3}}
\def\g_a#1#2#3{\Gamma^{#1}_{{#2}{#3}}}

\def\slash#1{\rlap{\kern0.07em/}#1}			
\def\lslash{\rlap{\kern-0.03em/}l}			
\def\slashcap#1{\rlap{\kern0.25em/}#1}			
\def\hta{\rlap{\kern-0.04em/}h}				

\def\vev#1{\left\langle{#1}\right\rangle}		
\def\ma2#1#2#3#4{\left(\matrix{&{#1}&{#2}\cr&{#3}&{#4}}\right)}
\def\ma3#1#2#3#4#5#6#7#8#9{\left(\matrix{&{#1}&{#2}&{#3}\cr&{#4}&{#5}&{#6}
    \cr&{#7}&{#8}&{#9}}\right)}

\def\tspino#1#2#3{\left(\begin{array}{c}{#1}\\{#2}\\{#3}\end{array}\right)}
\def\vettore#1#2#3#4#5#6{\left(\begin{array}{c}{#1}\\{#2}\\{#3}\\{#4
}\\{#5}\\{#6}\end{array}\right)}
\def\ga#1#2#3{\Gamma^{#1}_{{#2}{#3}}}

\newcommand{\bbox}[1] { {\bf #1} }

\def\bbox#1{%
\relax\ifmmode
\mathchoice
{{\hbox{\boldmath$\displaystyle#1$}}}%
{{\hbox{\boldmath$\textstyle#1$}}}%
{{\hbox{\boldmath$\scriptstyle#1$}}}%
{{\hbox{\boldmath$\scriptscriptstyle#1$}}}%
\else
\mbox{#1}%
\fi
}
\begin{document}



\pagenumbering{arabic}

\begin{titlepage} \vspace{0.2in} 
\begin{center} {\LARGE \bf  Gas of wormholes: a possible ground
state of Quantum Gravity
\\} \vspace*{0.3cm}
{\bf G.~Preparata, S.~Rovelli, S.-S.~Xue$^{(a)}$}\\ \vspace*{0.5cm}
Dipartimento di Fisica dell'Universit\`a and INFN - sezione di Milano, 
Via Celoria 16, Milan, Italy\\
(a) I.C.R.A. - International Center of Relativistic Astrophysics and INFN, La
Sapienza, 00185 Rome, Italy. \\ \vspace*{0.8cm}

{\bf   Abstract  \\ } \end{center} \indent

In order to gain insight into the possible Ground State of Quantized
Einstein's Gravity, we have derived a variational calculation of the
energy of the quantum gravitational field in an open space, as
measured by an asymptotic observer living in an asymptotically flat
space-time. We find that for Quantum Gravity (QG) it is energetically
favourable to perform its quantum fluctuations not upon flat
space-time but around a ``gas'' of wormholes of mass $m_p$, the Planck
mass ($m_p\simeq 10^{19}$GeV) and average distance $l_p$, the Planck
length $a_p$($a_p\simeq 10^{-33}$cm). As a result, assuming such
configuration to be a good approximation to the true Ground State of
Quantum Gravity, space-time, the arena of physical reality, turns out
to be well described by Wheeler's quantum foam and adequately modeled
by a space-time lattice with lattice constant $l_p$, the Planck
lattice. 

  \vspace*{1cm} \\
\end{titlepage}

\section{Introduction}
 
Among the fundamental interactions of Nature, since the monumental contribution
of Albert Einstein, Gravity plays the central role of determining the structure
of space-time, the arena of physical reality. As well known, in classical
physics a world without matter, the Vacuum, has the simplest of all structures,
it is flat (pseudoeuclidean); but in quantum physics? This is the focal
question that has occupied the best theoretical minds since it became
apparent, at the beginning of the 30's, that Quantum Field Theory (QFT') is
the indispensable intellectual tool for discovering the extremely subtle ways
in which the quantum world actually works. Thus the problem to solve was to
find in some way or other the Ground State (GS) of Quantum Gravity (QG), which
determines the dynamical behaviour of any physical system, through the
non-trivial structure that space-time acquires as a result of the quantum
fluctuations that in such state the gravitational field, like all quantum
fields, must experience. Of course this problem, at least in the
non-perturbative regime, is a formidable one, and many physicists, J.A.~Wheeler
foremost among them, could but speculate about the ways in which the expected
violent quantum fluctuations at the Planck distance $l_p$ ($l_p=10^{-33}cm$)
could change the space-time structure of the Vacuum, from its classical,
trivial (pseudoeuclidean) one. And Wheeler's conjecture, most imaginative and
intriguing, of a space-time foam vividly expressed the intuition that at the
Planck distance the fluctuations of the true QG ground state would end up in
submitting the classical continuum of events to a metamorphosis into an
essentially discontinuous, discrete structure\footnote{We should like to recall
here that, based on Wheeler's idea, a successful research program was initiated
a few years ago to explore the consequences of the Standard Model
($SU_c(3)\otimes SU_L(2)\otimes U_Y(1)$) in a discrete space-time, conveniently
modeled by a lattice of constant $l_p$, the Planck lattice (PL).} 

It is the purpose of this paper to give a detailed account of the 
results of an investigation on a possible QG ground state, which has 
been summarily reported in a recent letter\cite{letter}. 
The starting point of our attack
is the realization that QG can be looked at as a non-abelian gauge theory whose
gauge group is the Poincar\'e group. Following the analysis performed by one of
us (GP)\cite{gp} of another non-abelian gauge theory QCD (whose gauge group is
$SU_c(3)$), we decided to explore the possibility that the energy density (to
be appropriately defined, see below) of the quantum fluctuations of the
gravitational field around a non-trivial classical solution of the Einstein's
field equations for the matterless world, could be lower than the energy of the
perturbative ground state (PGS), which comprises the zero point fluctuations of
the gravitational field's modes around flat space-time. Indeed in QCD it was
found that the unstable modes (imaginary frequencies) of the gauge fields 
around the classical constant chromomagnetic field solution of the empty space
Yang-Mills equations, in the average screen completely the classical
chromomagnetic field, allowing the interaction energy between such field and
the short wave-length fluctuations of the quantized gauge field, to lower the
energy density of such configuration below the PGS energy density. Thus we
decided to try for QG the strategy that was successful in QCD, i.e. 

\begin{itemize}
\begin{enumerate} 

\item select a class of empty space classical solutions
of Einstein's equations that is simple and manageable; 

\item evaluate the spectrum of the small amplitude fluctuations of the 
gravitational field around such solutions; 

\item set up a variational calculation of the appropriately defined
energy density in the selected background fields; 

\item study the possible screening by the unstable modes (if any) of the
classical background fields. 

\end{enumerate} 
\end{itemize}

As for point (1) we have chosen the Schwarzschild's
wormhole-solutions\cite{s}, the simplest class of solutions of Einstein's
equation after flat space-time. In order to achieve (2) the
Regge-Wheeler\cite{rw} expansion has been systematically employed, yielding 
a well defined set of {\it unstable modes} (for S-wave). This important
result, already indicated in previous independent work \cite{d}, renders
the development of the points (3) and (4) both relevant and meaningful, the
former point yielding a lowering of the energy density due to the interaction 
of the short-wave length
modes with the background gravitational field, the latter exhibiting the
(approximate) cancellation of the independent components of the tensor of the
Schwarzschild's wormholes by the S-wave unstable modes. As a result flat
space-time, like the QCD perturbative ground state, becomes ``essentially
unstable'', in the sense that upon it no {\it stable} quantum dynamics can be
realized. On the other hand a well defined ``gas'' of wormholes appears as a
very good candidate for the classical configuration around which the quantized
modes of the gravitational field can {\it stably} fluctuate. But a discussion
of the physical implications of our findings must await a more detailed
description of our work, which we are now going to provide.

\section{The Schr\"odinger functional approach}

In order to develop a functional strategy aimed at determining the
Ground State of Quantum Gravity, which parallels the approach
developed for QCD \cite{gp}, we must first identify an appropriate
energy functional. In General Relativity this is a non-trivial problem
for, as is well known, in the canonical quantization procedure, first
envisaged by Dirac \cite{dirac} and Arnowitt, Deser and Misner
(ADM) \cite{a}, due to general covariance the local Hamiltonian is
constrained to annihilate the physical ground state, a fact that in the
Schr\"odinger functional approach is expressed by the celebrated
Wheeler-DeWitt equation \cite{dewitt}. However we note that  the
problem we wish to solve concerns  the minimization of the total
energy of an ``open space'', in which there exists a background metric
field that becomes  ``asymptotically flat'', i.e.~that for spatial
infinity  ($r = |\vec x| \rightarrow \infty$) behaves as
\begin{equation}
g_{\mu \nu} \longrightarrow \eta _{\mu \nu}^{(M)}
+ O({1\over r}),
\label{asy}
\end{equation}
where $\eta _{\mu \nu}^{(M)}$ is the Minkowski metric. In
the conventional canonical formulation, space-time is foliated into
spacelike slices $\Sigma$ with constant values of the time parameter $t$;
the asymptotic condition (\ref{asy}) determines the asymptotic behaviour of
the canonical variables: the spatial 3-metric $g_{ij}$ on $\Sigma$, 
the 
conjugate momenta $\pi^{ij}$, the ``lapse-function'' $N$ and the ``shift-
vector'' $N_i$ \cite{a} as:
\begin{eqnarray}
g_{ij} & \longrightarrow & \delta _{ij} + O(r^{-1})\nonumber\\
\pi^{ij} & \longrightarrow & O(r^{-2})\nonumber\\
N & \longrightarrow & 1 \nonumber\\
N_i  & \longrightarrow & 0.
\label{falloff}
\end{eqnarray}
Let us consider the ADM-energy \cite{a}, which in
cartesian coordinates is given by ($\partial\Sigma$ is the boundary of
$\Sigma$ , ``$,k$'' denotes partial derivative with respect to $x_k$, 
and $G=l_p^2$ is the Newton constant)
\begin{equation}
E_{ADM}={1\over16\pi G}\int_{\partial\Sigma}dS^k \delta^{ij}
(g_{ik,j}-g_{ij,k}).
\label{adm}
\end{equation}
We should like to point out that $E_{ADM}$ is just the energy that
an asymptotic observer attributes to a space region $\Sigma$ whose time 
foliation he is
keeping anchored to his (asymptotically) flat metric: the boundary
conditions select a privileged reference frame (up to Lorentz
transformations) that implicitly defines {\it the} physically relevant
energy. In particular, the asymptotic condition on $N$ fixes the
``boundary time'' unequivocally: the asymptotic observer is the only
possessor of an idealized clock that allows him to describe quantities
associated to the whole physical system without introducing
material clocks (i.e. auxiliary fields). In this sense, he is also the only
one that can really be termed as an idealized, non-interfering ``observer''
capable of describing geometry at the quantum level in terms of evolution,
not merely in terms of correlation between variables, thus giving a
full meaning to the expression ``quantum 
geometrodynamics''\cite{w}.

At the classical level, the definition of $E_{ADM}$ fixes
Minkowski geometry as the zero point of the total energy; the proof of its
positivity \cite{shy} can be looked at as the statement that flat space-time is
the (unique) vacuum of General Relativity. This explains why the
first steps towards the quantization of the theory have been based on a 
perturbative
approach on the flat background, with the fluctuating self-interacting 
field interpreted in the conventional particle view as creating and
annihilating {\it gravitons}, which propagate in pseudo-euclidean space: 
in this sense, we call flat
space-time filled with gravitons performing zero-point fluctuations 
the ``perturbative ground state'' (PGS).

We now know that the perturbative approach was doomed to fail: the 
non-renormalizability of the theory does not allow to make any 
meaningful and predictive perturbative expansion. On the other hand,
analyzing the theory to the lowest non-trivial order around a curved 
background may give us important indications of how the deadly 
``impasse'' of the perturbative approach may be finally overcome and 
give back to the simplest form of QG its status and role of a ``bona 
fide'' Quantum Field Theory. 

We thus study the quantum fluctuations of the gravitational field upon 
a generic, asymptotically flat {\it stationary} background geometry, 
solution of the sourceless Einstein's equations; in particular, for the 
background metric
we can choose a foliation orthogonal to Killing timelike vectors and
put it in static form. On a given slice $\Sigma$, the 3-metric is thus given
by
\begin{equation}
g_{ij}(x)=\eta_{ij}(x)+h_{ij}(x),
\label{h}
\end{equation}
where $\eta_{ij}(x)$ is the spatial background metric and $h_{ij}(x)$
the fluctuation to be quantized ($x \in \Sigma$). We can now expand
the total ( the sum of the Hamiltonian and $E_{ADM}$) energy
$E$ of space in powers of the fluctuations $h_{ij}(x)$ (the number 
$(n)$ 
denotes the order of the expansion):
\begin{equation}
E=E^{(0)}_{ADM}+E^{(1)}_{ADM}+\sum_{n\ge1}\int_\Sigma
d^3x (N{\cal H}^{(n)}+N_i{\cal H}^{i(n)}).
\label{energy}
\end{equation}
${\cal H}$ and ${\cal H}^i$ are the super-hamiltonian and super-
momentum, as defined by ADM \cite{a}:
\begin{eqnarray}
{\cal H} &\equiv& T+V\nonumber\\
T&=& {16\pi G} {\cal G}_{ijkl}\pi ^{ij}\pi ^{kl}\nonumber\\
V&=& -  \frac{g^{1/2}}{16\pi G}R
\label{superh}
\end{eqnarray}
where $g$ is the determinant of 3-metric and $R$ the corresponding
curvature scalar, with the ``supermetric'' ${\cal G}_{ijkl}$ given by
\begin{equation}
{\cal G}_{ijkl} \equiv \frac{1}{2}g^{-1/2}(g_{ik}g_{jl} +
g_{il}g_{jk} - g_{ij}g_{kl}),
\end{equation}
while ($;$ is the covariant derivative with respect to $g_{ij}$)
\begin{equation}
{\cal H}^i \equiv -2\pi ^{ij}_{~~;_j}.
\label{superm}
\end{equation}
We note that, a priori, also $N$ and $N_i$ can be expanded in
(\ref{energy}); while the background terms $N^{(0)}$ and
$N_i^{(0)}$ are fixed functions, subject to the asymptotic conditions
(\ref{falloff}), the higher order terms represent true fluctuations in
the lapse function and the shift vector: variations of $N$ and $N_i$ 
yield, 
at the classical level, the constraints:
\begin{eqnarray}
{\cal H} = \sum_{n\ge1} {\cal H}^{(n)} = 0\nonumber\\
{\cal H}^i = \sum_{n\ge1} {\cal H}^{i~(n)} = 0,
\label{vs}
\end{eqnarray}
leaving in the expression of the classical energy only the 
$E^{(\circ)}_{ADM}$ term.

Since $\eta_{ij}$, $N^{(0)}$ and $N_i^{(0)}$ form together a
solution $\eta_{\mu \nu}$ of  matterless Einstein's equations, the
linear term in the canonical Lagrangian density  must be a total divergence. In our case,
where $\eta_{\mu \nu}$ is  static, this is a purely spatial divergence
and, keeping the asymptotic flatness
of  background in mind, it must necessarily coincide with
$E^{(1)}_{ADM}$. Thus, we have:
\begin{equation}
E^{(1)}_{ADM}=
{1\over16\pi G}\int_{\partial\Sigma}dS^k (h_{kj,j}-h_{jj,k}) =
 -\int_\Sigma d^3x (N^{(0)}{\cal H}^{(1)}+
N_i^{(0)}{\cal H}^{i(1)}),
\label{adm1}
\end{equation}
that, together with the constraint (\ref{vs}) allows us to rewrite the
energy (\ref{energy}) as
\begin{equation}
E=E^{(0)}_{ADM}+\sum_{n\ge2}\int_\Sigma d^3x (N^{(0)}
{\cal H}^{(n)}+N_i^{(0)}{\cal H}^{i(n)}).
\label{adm2}
\end{equation}
Note the survival of only the classical $(0)$-order terms in $N$
and $N_i$.

The quantization of the theory promotes the canonical pair
$h_{ij}(x),\pi ^{kl}(y)$ on $\Sigma$ to operators obeying the
commutation rules:
\begin{equation}
[h_{ij}(x) , \pi ^{kl}(y)] = i\hbar \delta_i^{(k}
\delta_j^{l)} \delta^{(3)}(x - y),
\label{ccr}
\end{equation}
and acting on a Hilbert space of functionals $\Psi$ that are annihilated
by the constraints (\ref{vs}).
The evolution of the physical states $\Psi$ is governed by the
``Schr\"odinger equation"
\begin{equation}
i\hbar \partial_t\Psi = E\left[ h_{ij}, \pi^{kl}\right]\Psi,
\label{se}
\end{equation}
where the Hamiltonian operator is given by (\ref{adm2}). Note that this is just
the description of the quantum dynamics made by the asymptotic
observer at infinity.

We point out that our definition of  the Hilbert space is truly consistent
within our restriction of phase space to two pairs of canonical
operators, obtained by the gauge conditions (that respect (\ref{vs})).
This does not mean a loss of  invariance (and of physical reality) at all:
despite its look, the total energy (\ref{adm2}) is nothing but 
a rearrangement of the {\it invariant} ADM-energy. 

We should also be aware that the configuration space representation
of the canonical operators
\begin{eqnarray} 
h_{ij}(x) & \longrightarrow & h_{ij}(x) \times\nonumber \\
\pi ^{ij}(x) & \longrightarrow & 
-i\hbar \frac{\delta}{\delta h_{ij}(x)},
\label{rep}
\end{eqnarray}
acting on state functionals
\begin{equation}
\Psi \longrightarrow \Psi [h_{kl}](t)
\label{state}
\end{equation}
is not easily manageable beyond the 1-loop level, where connected ghost
terms appear. Beside that, for $n\ge3$ the expansion of the Hamiltonian
operator (\ref{adm2}) contains products of conjugate operators, 
thus posing an ordering problem. These problems are related to the bad 
ultraviolet divergences that would still yield a non-renormalizable 
behaviour, the possible solution of which emerges from the results  of 
Section 6, which show that the structure of the vacuum is, with good 
probability, essentially discontinuous at the Planck scale $l_p$. 
Thus, in the rest of our analysis the QFT we shall work with will be 
cut-off at the Planck scale, having clearly in mind that our results 
will only be meaningful if consistent with this fundamental 
assumption (see Section 6).

As for the constraints (\ref{vs}), the problems are easier to solve.
In fact we first notice that at the lowest order, the Hamiltonian
operator retains only quadratic terms in the fields, on which  we
have to impose consistently {\it first} order constraints, that do not annihilate the
quantum energy. The following terms in the expansions (\ref{vs})
can be enforced through a systematic correction of the state
functional $\Psi$ that readapts non-physical degrees of freedom
order by order, thus not affecting the dynamics based on the degrees 
of freedom
(two for each space point) isolated at the lowest level.
\footnote {See for example the procedure followed in Appendix B of
ref \cite{gp}.}
Thus in spite of the problems typical of Quantum Gravity, the parallelism
with the situation in QCD \cite{gp} is fully regained.

According to our fundamental assumption to cut the theory at the 
Planck scale, we shall perform a 1-loop calculation, with the
Hamiltonian operator truncated at $n=2$. We simply
adopt the representation (\ref{rep}) and (\ref{state}), thus obtaining the
Schr\"odinger equation
\begin{equation}
i\hbar \partial_t \Psi[h_{ij}](t) =
H[h_{ij} , -i\hbar \frac{\delta}{\delta h_{kl}}] \Psi[h_{ij}](t) ,
\label{se2}
\end{equation}
where
\begin{equation}
H = M + \int_{\Sigma} d^3xN^{(0)}
{\cal H}^{(2)}
(h_{ij}(x), -i\hbar \frac{\delta}{\delta h_{ij}(x)}),
\label{opH}
\end{equation}
with $\Psi[h_{ij}](t)$ annihilated by the first order constraints
\begin{eqnarray}
{\cal H}^{(1)}(h_{ij}(x), -i\hbar \frac{\delta}{\delta h_{ij}(x)}) 
\Psi[h_{ij}](t) = 0\nonumber\\
{\cal H}^{i~(1)} (h_{ij}(x), -i\hbar \frac{\delta}{\delta h_{ij}(x)})
\Psi[h_{ij}](t)= 0.
\label{qcs2}
\end{eqnarray}
Setting, as usual,
\begin{equation}
\Psi[h_{ij}](t) = e^{-iEt/\hbar}\Psi[h_{ij}],
\label{-iet}
\end{equation}
the problem can be reduced to the eigenvalue equation
\begin{equation}
H[h_{ij} , -i\hbar \frac{\delta}{\delta h_{ij}}] \Psi[h_{ij}] =
E \Psi[h_{ij}].
\label{eigen}
\end{equation}
We can now investigate the ground state of the theory. Instead of
solving directly the eigenvalue equation, we look for the minimization of
the expectation value of  $H$ on a class of gaussian wave-functionals:
\begin{equation}
E^{(2)} \equiv \int [{\cal D}h] \Psi^*[h_{ij}]
H\left[ h_{ij},-i\hbar {\delta \over \delta h_{ij}}\right]\Psi[h_{ij}].
\label{e2}
\end{equation}
If the background is stable under the action of quantum fluctuations, at
the 1-loop level this result coincides with the solution of  (\ref{eigen});
if, on the contrary, simple minimization  leads to an imaginary part in
$E^{(2)}$, then we have discovered an unstable configuration, whose 
physical meaning must be investigated. We demonstrate in the next
section that the latter case occurs when $h_{ij}$ fluctuate around
the ``wormhole solution'' $\eta_{ij}$ discovered by Schwarzschild in
1916 \cite{s}, whose line elements in polar coordinates are given by
($2GM < r < +\infty$)
\begin{equation}
ds^2=-{r-2GM\over r} dt^2 
+{r\over r-2GM}dr^2 +r^2(d\theta^2+\sin^2\theta d\phi^2)
\label{sm}
\end{equation}
and depend on the single parameter $M$, the ADM-mass, such that
\begin{equation}
E^{(0)}_{ADM}=M.
\label{adm3}
\end{equation}

\section{Quantum Fluctuations on a Schwarzschild Background }

We shall now address the problem to evaluate the expectation value on 
a gaussian trial functional of the Hamiltonian \reff{opH} where, 
according to our fundamental hypothesis (to be checked for consistency
at the end of the calculation), we keep only the quadratic terms in the 
field quantum fluctuations $h_{ij}$. This truncation corresponds to 
the one-loop approximation. From a classical standpoint this amounts 
to a calculation of the energy carried by the quantized 
gravitational waves propagating on a fixed background, in 
the weak field approximation.

In our analysis we shall follow closely the steps of the ref.\cite{gp}, 
where a similar calculation was carried out for a Yang-Mills theory.
We begin by constructing the Hilbert space of the states of the 
gravitational field, introducing the following scalar product:
\beq
\langle \langle \Psi |\Psi' \rangle \rangle 
\equiv \int({\cal D}h) \triangle _{FP} \Psi ^* [h_{ij}] \Psi'[h_{ij}],  
\label{3.1}
\eeq
where $({\cal D}h)$ denotes the measure of the functional space and 
$\triangle_{FP}$ represents the Fadeev-Popov determinant, depending 
on the gauge adopted, necessary to recuperate the gauge-invariance, 
i.e.~the general covariance of QG. The Hilbert space will thus be the
space of the state-vectors $\Psi[h_{ij}](t)$ that with the metric 
(\ref{3.1}) are normalizable. We note that for an infinitesimal 
coordinate transformation the quantum field $h_{ij}$ gets transformed 
as:
\beq
h_{ij}(x) \longrightarrow h_{ij}(x) - \xi _{i|j}(x) - \xi _{j|i}(x)  ,
\eeq
just like a weak classical field. And in our approximation, being 
the gauge-conditions (\ref{vs}) linear in the field $h_{ij}$, the 
determinant $\triangle_{FP}$ does not depend on $h_{ij}$ and can be
therefore neglected. For a generic operator $\hat{O}[h_{ij},\pi 
^{il}]$,
the expectation value on a state $\Psi[h_{ij}](t)$ can be defined as:
\beq
\langle \langle \hat{O} \rangle \rangle _{\Psi}(t) \equiv 
\frac{\langle \langle \Psi | \hat{O}\Psi \rangle \rangle}
{\langle \langle \Psi | \Psi \rangle \rangle}.
\eeq

Let us consider now a hypersurface $\Sigma$ at a fixed time $t$. We 
wish to compute the expectation value of the (truncated) Hamiltonian 
on 
the gaussian trial functional:
\beq\label{gauss}
\Psi_G[h_{ij}] = \exp~-\frac{1}{4}\int_{x,y} 
h_{ij}(\vec{x}) \Gamma ^{ijkl}(\vec{x},\vec{y}) h_{kl}(\vec{y})
\eeq
where $\int_x \equiv 
\int_{\Sigma}d^3\vec{x} \eta^{1/2}(\vec{x})$ and 
$\eta^{1\over2}(\vec{x})=\sqrt{r\over r-2m}$ ($m=MG$ is the 
one half Schwarzschild radius). In order to get a normalizable $\Psi_G[h_{ij}]$ 
we require that $\Gamma ^{ijkl}(\vec{x},\vec{y})$ be real and 
positive,
symmetric under the exchanges 
$i \leftrightarrow j ~,~ k \leftrightarrow l ~,~ ij,\vec{x} 
\leftrightarrow kl,\vec{y}$.

The second order Hamiltonian density is given by
\beq
{\cal H}^{(2)}(\vec{x}) = T^{(2)}(\vec{x}) + V^{(2)}(\vec{x})
\eeq
with
\beq\label{h2-t}
T^{(2)} =-16\pi G \eta^{-1/2}(\eta^{ik} \eta^{jl} 
- \frac{1}{2} \eta^{ij} \eta^{kl})
\frac{\delta ^2}{\delta h_{ij}(\vec{x})\delta h_{kl}(\vec{x})}
\eeq
\beq\label{p}
V^{(2)} = -\frac{1}{16\pi G} \eta^{1/2}
(R^{(2)} + \frac{1}{2}h^k_kR^{(1)}).
\eeq
We observe that for a Gaussian wave-functional the expectation value 
of two fields is given by:
\beq
\langle \langle h_{ij}(\vec{x})h_{kl}(\vec{y}) 
\rangle \rangle_{\Psi} = G_{ijkl}(\vec{x},\vec{y})
\label{ccc}
\eeq 
where $G_{ijkl}(\vec{x},\vec{y})$ satisfies the relationship:
\beq\label{gamm}
\int_z \Gamma ^{ijmn}(\vec{x},\vec{z}) G_{mnkl}(\vec{z},\vec{y}) = 
\eta^{-1/2}(\vec{y})\delta ^i_{(k} \delta ^j_{l)} 
\delta ^{(3)}(\vec{x} - \vec{y}).
\eeq
In this way one gets for the expectation value of the 
Hamiltonian\footnote{In order to clearly separate the classical from 
the quantum (one-loop) contributions for the rest of this Section we 
shall keep the Planck constant $\hbar$, instead of putting it equal to 
one, as done in the natural unit system.} 
\beq\label{asp_h}
\langle \langle H \rangle \rangle_{\Psi} = 
M + \hbar \int_x N(\vec{x})[\frac{\alpha}{4}
(\Gamma^{ij}_{~~ij}(\vec{x},\vec{x}) - 
\frac{1}{2}\Gamma^{i~j}_{~i~j}(\vec{x},\vec{x})) + 
\frac{1}{\alpha}\hat{O}^{ijkl}(\vec{x})G_{klij}(\vec{x},\vec{x})],
\eeq
where
\beq
\alpha \equiv 16\pi G\hbar = 16\pi l_P^2   ,
\eeq
and the differential operator $\hat{O}^{ijkl}(\vec{x})$ is defined by
\beq
\int_x NV^{(2)} = \frac{1}{\alpha}
\int_x 
N(\vec{x})h_{ij}(\vec{x})\hat{O}^{ijkl}(\vec{x})h_{kl}(\vec{x}),
\eeq
and represents the ``potential'' contribution to the quadratic 
Hamiltonian \reff{opH}. In order to guarantee the 
general covariance of our calculation, it is necessary to impose on 
the physical state the quantum constraints, which in our approximation 
are
\beq\label{vv1-a}
{\cal H}^{(1)}\Psi = (h^{ij}R^{(0)}_{ij} - h^{ij}_{~~|ij} 
- h^{i~~|j}_{i|j})\Psi = 0\\
\eeq
\beq\label{vv1-b}
{\cal H}^{i(1)}\Psi = \pi ^{ij}_{|j}\Psi = 0
\eeq
which are obeyed provided,
\beq
\nabla _j(\vec{x})\Gamma ^{ijkl}(\vec{x},\vec{y}) = 0 
\eeq
where $\nabla _j(\vec{x})$ denotes the covariant derivative with 
respect to the background field; and, fixing the lapse function and 
the shift vector as
\begin{equation}
N = N^{(0)} = \sqrt{1-\frac{2m}{r}} ~,~ N^i = N^{i(0)} = 0,
\end{equation}
we have 
\begin{equation}
\nabla _i (\frac{h^i_j}{N}) = 0 ~,~ h^k_k = 0  .
\end{equation}
By consistency with the trace condition $h^k_k = 0$ we must also 
impose $\pi ^k_k \Psi = 0$, which yields the further constraint:
\beq\label{tgam}
\Gamma _i^{~ikl}(\vec{x},\vec{y}) = 
\Gamma _{~~~k}^{ijk}(\vec{x},\vec{y}) = 0   .
\eeq
The elements of the Hilbert space of the physical modes of the 
gravitational field $h_{ij}$ are thus the symmetric tensors of rank 2
$\phi_{ij}$, defined in $\Sigma$, obeying the boundary conditions
\reff{falloff} and the gauge conditions:
\beq\label{gaugephi}
\nabla _i (\frac{\Phi ^i_j}{N}) = 0 ~,~ \Phi ^k_k = 0,
\eeq
normalizable with respect to the scalar product:
\beq\label{scp}
\langle \Phi|\Phi' \rangle \equiv 
\int_x N^{-1}(\vec{x}) \Phi^{*~ij}(\vec{x})\Phi'_{ij}(\vec{x}).
\eeq
In this way we may construct in our Hilbert space a complete 
orthogonal system, by making use of the spectral decomposition of 
the operator $\hat O^{ijkl}(\vec{x})$ or, better,. of the operator
$\hat Q^{ijkl}$ defined as:
\beq\label{oq}
\int_x NV^{(2)} = \frac{1}{\alpha}
\int_x N(\vec{x})h_{ij}(\vec{x})\hat{O}^{ijkl}(\vec{x})h_{kl}(\vec{x}) = 
\int_x 
N^{-1}(\vec{x})h_{ij}(\vec{x})\hat{Q}^{ijkl}(\vec{x})h_{kl}(\vec{x}),
\eeq 
where in order to go from $\hat O$ to $\hat Q$ a total divergence
has been added to the integrand, without changing the ``potential'' 
contribution to the Hamiltonian. Thus the operator $\hat Q^{ijkl}$ becomes
in our Hilbert space a self-adjoint hermitian operator, whose 
eigenfunctions ($\rho$ denotes a complete set of indices), 
\beq\label{spettroq}
\hat{Q}_{ij}^{~~kl}\Phi ^{(\rho)}_{kl} = \lambda(\rho) 
\Phi ^{(\rho)}_{ij}  ,
\eeq 
build the sought complete orthonormal
basis. In this basis the ``propagator'' $G_{ijkl}(\vec{x},\vec{y})$ 
has the simple form;
\beq\label{gmode}
G_{ijkl}(\vec{x},\vec{y}) = \sum_{\rho} 
\frac{1}{2f(\rho)}\Phi^{(\rho)*}_{ij}(\vec{x})\Phi^{(\rho)}_{kl}(\vec{y}),
\eeq
where $f(\rho)$ denotes a set of variational parameters to be 
determined by the minimization of the expectation value (\ref{asp_h}).

From \reff{gamm} and \reff{gmode} we obtain,
\beq\label{gammode} 
\Gamma ^{ijkl}(\vec{x},\vec{y}) = \sum_{\rho} 2f(\rho) 
N^{-1}(\vec{x}) N^{-1}(\vec{y}) 
\phi_{(\rho)}^{*~ij}(\vec{x})\phi_{(\rho)}^{kl}(\vec{y}).
\eeq
We may now easily compute the expectation value \reff{asp_h},
and obtain:
\beq\label{Hmode}
\langle \langle H \rangle \rangle_{\Psi} = 
M + \frac{\hbar}{2} \sum_{\rho}
(\alpha f(\rho) + \frac{\lambda(\rho)}{\alpha f(\rho)})  .
\eeq
and minimizing with respect to the variational function $f(\rho)$, 
i.e.~imposing
\beq\label{minimo}
\frac{\delta}{\delta f(\rho)}
\langle \langle H \rangle \rangle_{\Psi} = 0  ,
\eeq
we readily get
\beq\label{fmin}
f(\rho) = \alpha ^{-1}\sqrt{\lambda(\rho)}  .
\eeq
which inserted in \reff{Hmode} finally yields:
\beq\label{hofame}
E = M + \hbar \sum_{\rho}\sqrt{\lambda(\rho)}  .
\eeq
All the above makes sense if and only if
\beq
\lambda(\rho)>0,
\eeq
i.e.~the eigenvalue of the ``potential'' operator $\hat Q$ are 
positive definite. If, instead, for some $\rho$ $\lambda(\rho)\le 0$
the one-loop approximation, yielding imaginary contributions to the
energy of the ground state, breaks down, showing that the Perturbative 
Ground State (PGS, $M\rightarrow 0)$ is {\it essentially} unstable.

This is precisely the situation found in the study of $SU(n)$ 
Yang-Mills theories\cite{gp} where, going beyond the one-loop 
approximation, one could easily check that the modes belonging to the 
sector where $\lambda(\rho)\le 0$ did not contribute to the energy of 
the state terms of $O(\hbar)$\footnote{Like it happens, according to
\reff{hofame}, to the modes belonging to the ``stable sector'', for 
which $\lambda(\rho) >0$.}, but rather of $O(1)$, just like the 
classical term $M$. This ``promotion'' of a {\it quantum}  $O(\hbar)$
contribution to a {\bf classical} $O(1)$ one, can be understood when we 
realize that the amplitude $f(\rho)^{-{1\over2}}$ of the modes with 
$\lambda(\rho)\le 0$(the ``unstable modes'')[see Eqs.(\ref{ccc}) and 
(\ref{gmode})]
is only prevented from becoming infinite by the neglected positive terms 
of $O(\hbar^2)$. In this way $\alpha f(\rho)$ becomes 
$O({1\over\hbar})$ and the {\it negative} contribution from the ``unstable 
modes'' is just {\it classical}, i.e.~$O(1)$. 

In the calculation of 
Ref.\cite{gp} one could {\it explicitly} prove that this ``promoted'' 
quantum contribution {\it completely} screens the classical positive
term (such as, in our case $M$), thus realizing a ``vacuum'' state 
whose energy density is way 
that of below the PGS, which as a result becomes unstable at 
all space-time scales. In the case of QG the problem of going beyond 
the one-loop approximation is formidable, utterly beyond our present 
means of analysis, however, as shall be discussed below, to figure out 
 the contributions to the energy of the trial states of possible 
``unstable modes'' with $\lambda(\rho)\le 0$ appears reasonably doable.

In order to precede further we must first compute the operator $\hat{Q}^{ijkl}(\vec{x})$ and then diagonalize it, which we shall do 
next.
Defining ${\bar h}^{ij} \equiv g^{ij} - \eta^{ij}$, we have:
\beq 
\bar{h}^{ij} = - h^{ij} + h^i_l h^{lj} - h^i_l h^l_m h^{mj} + ... 
\eeq 
and for the perturbative calculations it is useful to introduce:
\beq
S^a_{mn} = \Gamma^a_{mn} - \Gamma^{a~(s)}_{mn} 
\eeq
which turns out to be tensor.

Observing that (``$;$'' denotes  the full covariant derivative,
while ``$|$'' is the covariant derivative with respect to the Schwarzschild 
background)
\beq
h_{ja;i} = h_{jk|i} - S^l_{im} h_{lj} - S^l_{ij} h_{lk},
\eeq
we may write the Riemann and the Ricci tensors as:
\beq
\left\{ \begin{array}{l}
R^a_{bcd} = R^{a~(s)}_{bcd} + S^a_{bd|c} - S^a_{bc|d} +S^a_{mc} 
S^m_{bd} - S^a_{dm} S^m_{bc} \nonumber\\ R_{bc} = R^{~~(s)}_{bc} 
+S^a_{bc|a} - S^a_{ab|c} +S^a_{am} S^m_{bc} - S^a_{lb} S^l_{ac}
\end{array} \right.
\eeq
furthermore:
\beq
S^i_{jk} = {1 \ov 2} g^{ik} \Big( h_{jr|k} + h_{rk|j} - h_{jk|r} 
\Big).
\eeq
We may now compute the curvature scalar ${R} = g^{bc} {R}_{bc}$, 
which for convenience we 
decompose as the sum of three terms:
\[ 
   A \equiv g^{bc} R^{(s)}_{bc}; ~~~~ B \equiv g^{bc}S^l_{bc|l} - 
   g^{bc} S^l_{lb|c}; ~~~~ C \equiv g^{bc} \Big( S^l_{lm} S^m_{bc} - 
   S^l_{mb} S^m_{lc} \Big) .
\]
In this way to first order in $h_{ij}$ we have:
\beq
A^{(1)} = {m \ov r^3} \Big( - h + 3 h^1_1 \Big)
\eeq
while to second order we obtain: 
\beq
A^{(2)} = {m \ov r^3} \Big( h^a_i h^i_a - 3 h^1_i h^i_1 \Big).
\eeq
As for $B$ we have
\beq
B^{(1)} = \eta^{lm} S^{a(1)}_{lm|a} - \eta^{lm} S^{a(1)}_{la|m} = 
h^{ab}_{~~|ba} - h^{~a}_{|~a} 
\eeq
and
\begin{eqnarray}
B^{(2)} \!& = \Big\{ - \Big[ h^{ar} h^{~~|s}_{rs} \Big]_{|a} + {1\ov 
2} h^{ar}_{~~|a} h_{|r} - h^{lm} h^a_{m|al} \Big\} + \Big\{ h^{lm} h_{|m} 
- 2 h A^{(1)} \Big\} &\! \nonumber\\ \!& - 3A^{(2)} + h^{lm} 
h^{~~|a}_{lm~a} + {1 \ov 2} h^{lm}_{~~|r} h^{~~|r}_{lm} &\!
\end{eqnarray}
respectively. While for the third term one has:
\beq
C^{(2)} = {1 \ov 2} h_{|b} h^{ba}_{~|a} - {1 \ov 4} h_{|b} h^{|b} - 
{1 \ov 4} \Big( 2h^{al|b} h_{ab|l} - h^{am|b} h_{am|b} \Big) .
\eeq
By summing the different terms we obtain the following expansion of 
the scalar curvature:
\beq
R^{(0)} = 0
\eeq
\beq
R^{(1)} = \Big\{ h^{ab}_{~~|ba} \Big\} + \lt\{ - {m \ov r^3} h - 
h^{~a}_{|a} \ri\} + \lt\{ 3 {m \ov r^3} h^1_1 \ri\} 
\eeq
\begin{eqnarray}
\lefteqn{ R^{(2)} = \lt\{ - \Big[ h^{ar} h^{~~|s}_{rs} \Big]_{|a} + 
h^{ar}_{~|a} h_{|r} - h^{lm} h^a_{m|al}  \ri\} } \nonumber\\& & + 
\lt\{ h^{lm} h_{|lm} + 2 (h)^2 
{m \ov r^3} - 6{m \ov r^3} h h^1_1 - {1 \ov 4} h_{|b} h^{|b} \ri\} 
\nonumber\\& & + \lt\{ - {2m \ov r^3} \Big( h^a_i h^i_a - 3 h^1_i 
h^i_1 \Big) + h^{lm} h^{~~|a}_{lma} \ri. \nonumber\\& & \lt. + {3 \ov 
4} h^{lm}_{~~|r} h^{~~|r}_{lm} - {1 \ov 2} h^{al}_{~|b} h^b_{a|l} 
\ri\}.
\end{eqnarray}
But we are not done yet, we must expand the square root of the 
determinant of the metric $\sqrt{\| g \|}$.
Due to $R^{(0)} = 0$ we may stop at the first order, obtaining:
\begin{eqnarray}
\sqrt{ { \| {\bbox{g}} \| \ov \| {\bbox{\eta}} \| }} = 1 
+ {1 \ov 2} h^i_i + O(h^2.) 
\end{eqnarray}
In this way for $V^{(2)}$ we get the following expansion:
\begin{eqnarray}\label{potenziale}
\lefteqn{ V^{(2)} = - {1 \ov 16 \pi G} \sqrt{\| {\bbox{\eta}} \| } \Big[ {1 \ov 
2} h^{ab}_{~~|ba} \cdot h - {1 \ov 2} {m \ov r^3} (h)^2 - {1 \ov 2} 
h^{~~a}_{|a} } \nonumber\\& & + 3 {m \ov r^3} h^1_1 h - \Big( h^{ar} 
h^{~~|s}_{rs} \Big)_{|a} + h^{ar}_{~~|a} h_{|r} - h^{lm} h^a_{m|al} + 
h^{lm} h_{|lm} \nonumber\\& & + 2 (h)^2 {m \ov r^3} - 6 {m \ov r^3} h h^1_1 
- {1 \ov 4} h_{|b} h^{|b} - {2m \ov r^3} \Big( h^a_i h^i_a - 3 h^1_i 
h^i_1 \Big) \nonumber\\& & + h^{lm} h^{~~|a}_{lm~~a} + {3 \ov 4} h^{lm}_{~~|r} 
h^{~~|r}_{lm} - {1 \ov 2} h^{al}_{~~|b} h^b_{a|l} \Big] .
\end{eqnarray} 
We are now ready to compute the operator $\hat Q^{ijkl}$. Let us fix the 
gauge as: $h = 0$ and $\lt( {h^{ij} \ov N} \ri)_{|j} = 0$,
the latter condition by use of the former giving
\beq
h^j_{i|j} = {m \ov r^2} h_{1i}. 
\eeq
We also note that given a vector $T^m$, one has
\beq
\int_x N [T^m]_{|m} = - \int_x N_{,m} T^m = - \int_x T^1 {1 \ov \sqrt{1 - 
{2m \ov r}}} {m \ov r^2} = \int_x T^1 N \Gamma^1_{11} .
\eeq
Thus we obtain for the integral of the second order potential,
\begin{eqnarray}
& & \int_x N V^{(2)} = \int_x N {1 \ov 4} h^a_b \lt( - \De^2 
h^b_a + 3 
h^b_{a|1} {m \ov r^2} - h^{~~|b}_{a|} {m \ov r^2} - h^b_{1|a} {m \ov 
r^2} \ri. \nonumber\\& & \mbox{} ~~ \lt. + h^b_c R^c_a + h^c_a R^b_c 
- {8 m^2 \ov r^3 (r - 2m)} \delta^1_{(a} h^{b)}_1 \ri).
\end{eqnarray}
Recalling Eq.(\ref{oq}), Eq.(\ref{spettroq}) becomes:
\begin{eqnarray}
& & N^2 \lt[ - \De^2 \phi^b_a + 3 \phi^b_{a|1} {m \ov r^2} - 
\phi^{~~|b}_{a1} {m \ov r^2} - \phi^b_{1|a} {m \ov r^2} + \phi^b_c 
R^c_a + \phi^c_a R^b_c \ri. \nonumber\\& & \mbox{} ~~ \lt. - {8 m^2 \ov 
r^3 (r - 2m)} \delta^1_{(a} h^{b)}_1 \ri] + v^b_{|a} + v^{~|b}_a + 
\eta^{~b}_a \tau = 4 \lambda \phi^b_a,
\label{ee}
\end{eqnarray}
where we have added the terms in $v^i$ e $\tau$ in order to remain inside the Hilbert space of the tensors 
$\phi^{ij}$,
which obey the conditions [see Eq.(\ref{gaugephi})]
\begin{eqnarray}
\lefteqn{ \phantom{pippi} \phi^i_i = 0 } \nonumber\\& & \phi ^j_{i~|j} 
= {m \ov r^2} \phi_{1i}.
\label{effe}
\end{eqnarray} 
By taking the trace of eq.(\ref{ee}) with $\eta^a_b$ we get immediately 
that
\beq
\tau = - {2 \ov 3} v^b_{~|b} - N^2 \lt( R^1_1 \phi^1_1 - {10 m^2 \ov 3 
r^3} {\phi^1_1 \ov r - 2m} \ri).
\label{gi}
\eeq
By consistency, we must require that the terms in $v_i$ do not 
contribute to the potential energy, this obviously implies that 
\beq
\int_x \Big[ N^{-1} 
\phi^{ij}_{(\rho )} v^{(\rho )}_i \Big]_{|j}=0,
\eeq
which requires that for $r\rightarrow\infty$ the following conditions 
be satisfied:
\begin{eqnarray}
v_1 &\rightarrow& O(r^{-1}) \nonumber\\ 
v_2, v_3 &\rightarrow& O(1),
\label{cre}
\end{eqnarray}
and 
\beq
\lim_{r \rightarrow + \infty} r^2 \int d\Omega \phi^i_1 v_i = \lim_{r 
\rightarrow 2m} r^2 \int d\Omega \phi^i_1 v_i =0.
\label{pa}
\eeq

\section{The eigenvalues and eigenmodes of the second order potential}

Due to the spherical symmetry of the Schwarzschild background a particularly 
suitable method to obtain the solutions of the eigenvalue problem 
posed by Eq.\reff{ee} is the one devised by T.~Regge and J.A.~Wheeler
\cite{rw}, for the study of the small (classical ) fluctuations around 
the Schwarzschild solution. By making use of this method the eigenfunction are 
separated in two classes (see appendix B), the ``even'' solutions 
with parity $(-1)^l$, equal to the parity of spherical harmonics 
$Y_{lm}(\theta,\phi)$, and the ``odd'' solutions with opposite 
parity.For the ``even'' solutions, if we set
\beq
\left\{ \begin{array}{l}
\phi^1_1 = H (r) Y_{lm} (\theta  , \varphi ) \nonumber\\ 
\phi^2_2 = \Big( G_1 (r) + G_2 (r) \de^2_{\theta } \Big) Y_{lm} (\theta  , 
\varphi ) \nonumber\\ 
\phi^3_3 = \lt( G_1 (r) + {G_2(r) \ov \sin^2 \theta  } \de^2_\varphi + 
G_2(r) \cot \theta  \de_{\theta } \ri) Y_{lm} (\theta  , \varphi ) \nonumber\\
\phi^2_1 = K (r) \de_\theta  Y_{lm} (\theta  , \varphi ) \nonumber\\ 
\phi^3_1 = {K (r) \ov \sin^2 \theta  } \de_\varphi Y_{lm} (\theta  ,\varphi ) 
\nonumber\\ 
\phi^3_2 = {G_2(r) \ov \sin^2 \theta  } ( \de_\theta  - \cot \theta  ) 
\de_\varphi 
Y_{lm} (\theta  ,\varphi ),
\end{array} \right.
\label{comen}
\eeq
the eigenvalue equations will turn out to be, as we shall see in a 
moment, completely factorized. As for the vector $v_i$ \reff{ee}, factorization 
is achieved if we set,
\beq
\left\{ \begin{array}{l}
v_1 = U(r) Y_{lm}(\theta  ,\varphi ) \nonumber\\ 
v_2 = V(r) \de_\theta  Y_{lm}(\theta  ,\varphi ) \nonumber\\ 
v_3 = V(r) \de_\varphi Y_{lm}(\theta  ,\varphi ) 
 \end{array} \right.
\label{tddt}
\eeq
Substituting (\ref{comen}) and (\ref{tddt}) in (\ref{ee}) (see appendix C) 
we obtain a 
system of differential equations for the radial functions only:
\beq
\left\{ \begin{array}{l}
N^2  \left\{ \left[ {r -2m \over r} \de^2_r + 2{r -2m \over r^2} 
\de_r - {2\over 3} {6 r^2 - 27 mr + 23 m^2 \over r^3 ( r - 2m)} \ri ] 
H(r) \ri. \nonumber\\ ~~ + \left. 4 {r - 2m^{ } \over r^2} l(l+1) K(r) + 2 
{r -2m \over r^3} \Big( 2G_1 (r) - l(l+1) G_2(r) \Big) \ri\} 
\nonumber\\ ~~ = - H(r) 4 \lambda + {4 \over 3} \left[ {r - 2m \over r} 
\de_r U(r) - {r - 3m \over r^2} U(r) + {V(r)\over 2} {l (l+1) \over 
r^2} \ri] \nonumber\\ ~~~~ + N^2 {l(l+1) \over r^2} H(r)\\
\\
N^2 \lt\{ \lt[ {r -2m \over r} \de^2_r + 2{r -3m \over r^2} 
\de_r +{2 \over r^2} \ri] G_2(r) + 2 {2r - 3m \over r^2} K(r) \ri\} 
\nonumber\\ = - 4 \lambda G_2(r) + {2V(r) \over r^2 } + N^2 {\lp \over 
r^2} G_2(r) \\
\\
N^2 \lt\{ \lt[ {r -2m \over r} \de^2_r + 2{r -3m \over r^2} 
\de_r -{2 \over r^2} \ri] G_1(r) + {2\over r^2} G_2(r) \lp \ri. 
\nonumber\\ + \lt. \phantom{\lt[{r\ov r} \ri]} {2\over 3} {3r^2 - 
12 mr + 7 m^2 \over r^3 (r - 2m)} H(r) \ri\} \nonumber\\ = - 4 \lambda 
G_1(r) - {2 \over 3} \lt[ {r - 2m \over r} \de_r U(r) - {r - 3m \over 
r^2} U(r) - {V(r)\over r^2} l (l+1) \ri] \nonumber\\ + N^2 {l(l+1) 
\over r^2} G_1(r) \\
\\
N^2 \lt\{ \lt[ {r -2m \over r} \de^2_r + {4r -7m \over r^2} 
\de_r -{2r^2 -mr \cdot 10 + 10 m^2 \over r^3 (r -2m)} \ri] K(r) \ri. 
\nonumber\\ + \lt. {2r -3m \over r^3 ( r - 2m)} H(r) - {2\over r^3} 
\lt[ G_1(r) + \lt(1^{} - \lp \ri) G_2(r) \ri] \ri\} \nonumber\\ = - 
4 \lambda K(r) + {1\over r^2} \lt( U(r) + \de_r V(r) - {2\over r} V(r) \ri) 
+ N^2 {l(l+1) \over r^2} K(r) .
\end{array} \right.
\label{10}
\eeq
And by substituting the same expansions in the constraint 
Eq.(\ref{gaugephi}), we 
obtain,
\beq 
2 G_1(r) - G_2(r) \lp + H(r) = 0
\label{gau}
\eeq
\beq
\de_r H(r) + {3H(r) \over r} - {m \over r(r - 2m)} H(r) - K(r) \lp =0
\label{rigau}
\eeq
\begin{eqnarray}
\lefteqn{ \de_r K(r) r(r - 2m) + K(r) (4r - 8m) } \nonumber\\& & + 
G_1(r) +G_2(r) \lt[ 1 - \lp \ri]  = 0.
\label{qwer}
\end{eqnarray}

The boundary conditions \reff{cre} for $r\rightarrow\infty$ become:
\beq
\left\{ \begin{array}{l}
U(r) \rightarrow O \lt( {1 \ov r} \ri) \nonumber\\
\\
V(r) \rightarrow O (1) 
\end{array} \right.
\label{labeel}
\eeq
while \reff{pa} imply:
\begin{eqnarray}
& & \Big\{ r^2 \Big[ H(r) \cdot U(r) + \lp K(r) V(r) \Big] \Big\} 
\mid_{r = 2m} \nonumber\\& & \mbox{} ~~~~ ~~ = \Big\{ r^2 \Big[ H(r) 
\cdot U(r) + \lp K(r) V(r) \Big] \Big\} \mid_{r = + \infty} = 0.
\end{eqnarray}
Proceeding in a completely analogous way, for the ``odd'' solutions 
one gets:
\beq
\left\{ \begin{array}{l}
\phi^1_1 = 0 \nonumber\\
\\
\phi^2_2 = F_2 (r) {1 \ov \sin \theta  } [ \de_\theta - \cot \theta ] 
\de_\varphi Y_{lm} (\theta , \varphi) \nonumber\\ 
\\
\phi^3_3 = - \phi^2_2 \nonumber\\ 
\\
\phi^2_1 = - {F_1 (r) \ov \sin \theta } \de_\varphi Y_{lm} (\theta , 
\varphi ) \nonumber\\ 
\\
\phi^3_1 = {F_1 (r) \ov \sin \theta } \de_\theta  Y_{lm} (\theta  
,\varphi ) \nonumber\\ 
\\
\phi^3_2 = {1 \ov 2} {F_2 (r) \ov \sin \theta } \Big[ {1 \ov \sin^2 
\theta } \de^2_\varphi + \cot \theta - \de^2_\theta \Big] Y_{lm} (\theta 
, \varphi )
\end{array} \right.
\label{chimj}
\eeq
and
\beq
\left. \begin{array}{l}
v_1 = 0 \nonumber\\
v_2 = D(r) \lt( - {1 \ov \sin \theta  } \ri) \de_\varphi Y_{lm} (\theta , 
\varphi) \nonumber\\ 
v_3 = D(r) \sin \theta  \de_\theta  Y_{lm} (\theta  ,\varphi ).
\end{array} \right.
\eeq

And substituting these last expressions in the eigenvalue equations, 
we get the following relationships:
\begin{eqnarray}
& & N^2 \lt[ - {r - 2m \ov r} \de^2_r - {6r - 11m \ov r^2} \de_r + {\lp 
\ov r^2} + {2 \ov r^3} \lt( -3( r - 2m) \ri. \ri. \nonumber\\& & 
\mbox{} ~~ \lt. \lt. - {r - m \ov r - 2m} m \ri) \ri] F_1(r) = 4 
\lambda F_1(r) + {1 \ov r^2} \lt( \de_r - {2 \ov r} \ri) D(r) 
\label{aval}
\end{eqnarray}
and
\begin{eqnarray}
& & N^2 \lt[ \lt( - {r - 2m \ov r} \de^2_r - 2 {r - 3m \ov r^2} 
\de_r + {\lp 
\ov r^2} - {2 \ov r^2} \ri) F_2(r) \ri. \nonumber\\& & \mbox{} ~~~~ \lt. 
+ 2 {2r - 3m \ov r^2} F_1(r) \ri] = 4 \lambda F_2(r) - {1 \ov r^2} 
D(r). 
\label{raval}
\end{eqnarray}

Noting that the fluctuations' tensor automatically satisfies the 
traceless condition, we find that the gauge conditions yield
\beq
(r - 2m) (r \de_r + 4) F_1(r) - \lt( 1 - {\lp \ov 2} \ri) F_2(r) = 0
\label{vigau}
\eeq
as the only constraint.

The conditions \reff{cre} reduce to $D(r) \rightarrow O(1)$,
while \reff{pa} yield
\beq
\Big[ \lp r^2 F_1 (r) D(r) \Big] \mid_{r = 2m} = \Big[ \lp r^2 F_1 (r) 
D(r) \Big] \mid_{r = + \infty} = 0.
\eeq

An important observation is that for S-waves ($l=0$) the gauge 
conditions are sufficient to determine the form of the solution. 
In the other cases we shall solve the eigenvalue problem in the 
WKB approximation.

Setting now $l=0$ in our equation we realize at once that the ``odd'' 
part vanishes identically, while the first of the \reff{10} has a 
particularly simple form, depending on 
$H(r)$ and $G(r)$ only, the latter obeying through \reff{gau}
$H(r) = - 2 G_1 (r)$. The fluctuations' tensor acquires thus the 
following simple form:
\beq
\phi^a_b (r, \theta  ,\varphi ) = {1 \ov \sqrt{4 \pi}} \Big( 3 \delta^a_1 
\delta^1_b - \delta^a_b \Big) {1 \ov 2} H(r),
\eeq
where $H(r)$ is determined via eq.\reff{rigau}
\beq
\de_r H (r) + H (r) \lt[ {3 \ov r} + {1 \ov 2} \lt( {1 \ov r} - {1 \ov r 
- 2m} \ri) \ri] = 0,
\eeq
whose solution is
\[ 
   H(r) = A \sqrt{ 1 - {2m \ov r}} {1 \ov r^3},
\]
where the constant $A$ is determined by the normalization condition.
\beq
1 = \vev{\phi | \phi } = \int_\Sigma d^3 \xv \| {\bbox \eta} 
\|^{1/2} {1 \ov N} {1 \ov 4\pi} {3 \ov 2} H^2(r) = {A^2 \ov 16 m^3} 
\eeq
which yields $A = 4m^{3/2}$, so that
\beq
   H(r) = 4 m^{3/2} \sqrt{1 - {2m \ov r}} {1 \ov r^3}. 
\label{nellaprima}
\eeq
In Appendix D, we develop the calculation for the eigenfunctions' 
normalization in general.

It remains to be verified whether there exists an eigenvalue 
corresponding to \reff{nellaprima}, in other words, we are looking for 
a value $\lambda$ for which \reff{nellaprima} is a solution of the 
eigenvalue equations satisfying the boundary conditions
\reff{cre} and \reff{pa}. 
On substituting \reff{nellaprima} in the first of  
\reff{10}, and observing that
\[ 
   \de_r H(r) = - {3r - 7m \ov r (r - 2m)} H(r) 
\]
\[ 
   \de^2_r H(r) = {12r^2 - 56mr + 63 m^2 \ov r^2 (r- 2m)^2} H(r) 
\]
and
\[ 
   G_1(r) = - {1 \ov 2} H(r) 
\]
we find:
\begin{eqnarray}
& & {4 \ov 3} \lt( {r - 2m \ov r} {d \ov dr} - {r - 3m \ov r^2} \ri) U(r) 
\nonumber\\& & \mbox{} ~~ 
= \lt( 4 \lambda - {m (12r - 35m ) \ov 3r^4} \ri) 4m^{3/2} \sqrt{1 - {2m 
\ov r}}{1 \ov r^3}
\end{eqnarray}
whose solution is:
\beq
U(r) = \sqrt{{r \ov r - 2m}} \lt[ rC + 4m^{3/2} \lt( - {\lambda \ov r^2} + 
{m \ov 2r^5} - {5 \ov 4}{m^2 \ov r^6} \ri) \ri], 
\label{unstable}
\eeq
$C$ being an integration constant. Due to \reff{cre} $C$ must vanish, 
while \reff{pa} finally fixes the eigenvalue $\lambda$. Indeed from
\[ 
   \lim_{r \rightarrow 2m} r^2 H(r) U(r) = 0 
\]
one gets for the eigenvalue $\lambda$:
\beq
\lambda = - {1 \ov 64m^2} = - {1 \ov 64 (GM)^2}
\label{e}
\eeq

This finding of a {\it negative} eigenvalue $\lambda$, i.e.~of an 
``unstable mode''
around a ``wormhole'', in view of the discussion in section 3,
is a most significant result of our analysis that strengthens the 
analogy of the present calculation with the QCD one\cite{gp}. We must, 
however, point out that (\ref{e}) is not at all unexpected, since in 
ref.\cite{d} a similar finding has been reported in a somewhat 
different context.

As for the remaining stable modes we shall solve the eigenvalue 
problem in an approximate way, by the WKB method. We shall thus obtain a 
good description in the semiclassical region. Let us then write our 
solution in the form
\begin{eqnarray}
& \vettore{H(r) }{K(r) }{G_1 (r)}{G_2 (r)}{U(r) }{V(r) } & = \vettore{h(r) 
}{k (r)}{g_1 (r)}{g_2 (r)}{u(r) }{v(r) } e^{{i \ov \hbar } 
S(r)} \nonumber\\& & \mbox{} \simeq \vettore{h_0 (r) + \lt( {\hbar \ov 
i} \ri) h_1 (r) + \lt( {\hbar \ov i} \ri)^2 h_2 (r) + \cdot \cdot \cdot 
}{ k_0 (r) + \lt( {\hbar \ov i} \ri) k_1 (r) + \lt( {\hbar 
\ov i} \ri)^2 k_2 (r) + \cdot \cdot \cdot }{{g_1}_0 (r) + \lt( 
{\hbar \ov i} \ri) {g_1}_1 (r) + \lt( {\hbar \ov i} \ri)^2 {g_1}_2 (r) 
+ \cdot \cdot \cdot }{{g_2}_0 (r) + \lt( {\hbar \ov i} \ri) {g_2}_1 (r) 
+ \lt( {\hbar \ov i} \ri)^2 {g_2}_2 (r) + \cdot \cdot \cdot }{u_0 (r) + 
\lt( {\hbar \ov i} \ri) u_1 (r) + \lt( {\hbar \ov i} \ri)^2 u_2 (r) + 
\cdot \cdot \cdot }{v_0 (r) + \lt( {\hbar \ov i} \ri) v_1 (r) + \lt( 
{\hbar \ov i} \ri)^2 v_2 (r) + \cdot \cdot \cdot } \nonumber\\& & 
\mbox{} ~~~~ ~~~~ \cdot e^{{i \ov \hbar} \lt( 
s_0 (r) + {\hbar \ov i} s_1 (r) + \lt( {\hbar \ov i} \ri)^2 s_2 (r) + 
\cdot \cdot \cdot \ri) }
\end{eqnarray}

In order to simplify the analysis we enclose our system in a spherical box 
of radius $R\gg m$, and impose the condition that the field vanishes at the 
boundary, in 
this way the energy spectrum becomes discrete. By using the gauge 
constraints and equating the terms of the same order
$\displaystyle{\hbar \ov i}$ we have
\begin{eqnarray}
& & \vettore{H (r)}{K (r)}{G_1 (r)}{G_2 (r)}{U (r)}{V (r)} = \vettore{\lt( 
{\hbar \ov i} \ri)^{2} h_2 (r)}{\lt( {\hbar \ov i} \ri) \lt[ 1 - {\lp 
\ov 2} \ri] {{g_2}_0 (r) \ov {\dot{s}}_0 (r)} {1 \ov N^2 r^2} + \lt( 
{\hbar \ov i} \ri)^{2} {\cal H}_2 (r)}{{\lp \ov 2} {g_2}_0 (r) + \lt( 
{\hbar \ov i} \ri) {\lp \ov 2} {g_2}_1 (r) + \lt( {\hbar \ov i} \ri)^{2} 
\lt[ {\lp \ov 2} {g_2}_2 (r) - h_2 (r) \ri] }{{g_2}_0 (r) + \lt( {\hbar 
\ov i} \ri) {g_2}_1 (r) + \lt( {\hbar \ov i} \ri)^{2} {g_2}_2 (r)}{ 
u_0 (r) + \lt( {\hbar \ov i} \ri) u_1 (r) + \lt( {\hbar \ov i} \ri)^{2} 
u_2 (r)}{v_0 (r) + \lt( {\hbar \ov i} \ri)^{2} v_1 (r) + \lt( {\hbar 
\ov i} \ri)^{2} v_2 (r)} \nonumber\\& & \mbox{} ~~~~ ~~ e^{{i \ov \hbar 
} \lt( s_0 (r) + \lt( {\hbar \ov 
i} \ri) s_1 (r) + \lt( {\hbar \ov i} \ri)^{2} s_2 (r) \ri) },
\end{eqnarray}
with the conditions ($\dot{f}={d \ov dr} f (r)$):
\beq
\left. \begin{array}{l} 
h_2 (r) = \lp  {k_1 (r) \ov {\dot{s}}_0 (r)}  \nonumber\\
k_2 (r) = - {1 \ov {\dot{s}}_0 (r) } \lt[ {\dot{\cal H}}_1 (r) + {4 \ov 
r} k_1 (r) + {g_2}_1 (r) \lt( 1 - {\lp \ov 2} \ri) + k_1 {\dot{s}}_1 
(r) \ri] \nonumber\\ 
k_1 (r) = \lt[ 1 - {\lp \ov 2} \ri] {{g_2}_0 \ov {\dot{s}}_0 
(r)} {1 \ov N^2 r^2}.
\end{array} \right.
\label{nonsca}
\eeq
Inserting these expressions in the eigenvalue equations
\reff{10}, we can check that the following is a solution:
\begin{eqnarray}
\vettore{H (r)}{K (r)}{G_1 (r)}{G_2 (r)}{U (r)}{V (r)} & = & 
\vettore{\pm \lt( {\hbar \ov i } \ri)^2 \lp \lt[ 1 - {\lp \ov 2} \ri] 
{N^2 \ov 4 \lambda r^2}}{ \pm \lt( {\hbar \ov i } \ri) \lt[ 1 - {\lp 
\ov 2} \ri] {1 \ov 2 \sqrt{\lambda}r^2} - \lt( {\hbar \ov i } \ri)^2 
\lt[ 1 - {\lp \ov 2} \ri] {1 \ov 4 \lambda r^3}}{{\lp \ov 2} \mp \lt( 
{\hbar \ov i } \ri)^2 \lp \lt[ 1 - {\lp \ov 2} \ri] {N^2 \ov 4 \lambda 
r^2}}{1}{0}{0} \cdot \nonumber\\& & \mbox{} ~~~~ \cdot N^2 {c \ov r} 
e^{\pm {i \ov \hbar} 2 \sqrt{\lambda} r^* + \alpha (l) + O(\hbar)}
\label{tetd}
\end{eqnarray}
where $c$ is an integration constant, depending in general on $l$, and 
$r^* = r + 2m {\hbox {log}} \Big( {r \ov 2m} -1\Big)$. 
The constant $\alpha (l)$ is crucial for determining the eigenvalues.
As illustrated in Appendix E, $\alpha(l)$ can be determined by 
comparing (\ref{tetd}) with the exact solutions for the flat case 
$M=0$, obtaining 
\[ 
   \alpha (l) = -i (l + 1) {\pi \ov 2}. 
\]
We shall now determine the regime of validity of our approximation. We 
observe that 
\reff{tetd} has been obtained by expanding the eigenvalue equations up 
to the order $\Big( {\hbar \ov i} \Big)$. By considering the next 
order one gets:
\beq
s_2 (r) = \pm {1 \ov 4 \sqrt{\lambda}} \lt[ {1 \ov r} + {1 \ov 2m} \ln 
{r \ov r - 2m} \ri] .
\eeq
so that $\hbar |s_2 (r)| \ll 1$ if $2{\sqrt{\lambda} r 
\ov \hbar} \gg 1$, implying that, as expected, the approximation is 
good in the very high energy region. Before imposing the boundary 
conditions, let us analyse the behaviour of the ``odd'' part.

Looking for a solution of the form:
\begin{eqnarray}
\tspino{F_1 (r)}{F_2 (r)}{D(r)} & = & \tspino{L(r)}{M(r)}{d(r)} e^{{i 
\ov \hbar} A(r)} \nonumber\\& = & \mbox{} \tspino{L_0 (r) + {\hbar \ov 
i} L_1 (r) + \lt( {\hbar \ov i} \ri)^2 L_2 (r)}{M_0 (r) + {\hbar \ov 
i} M_1 (r) + \lt( {\hbar \ov i} \ri)^2 M_2 (r)}{d_0 (r) + {\hbar \ov 
i} d_1 (r) + \lt( {\hbar \ov i} \ri)^2 d_2 (r)} \nonumber\\& & \mbox{} 
\cdot e^{{i \ov \hbar} A_0 (r) + A_1 (r) + {\hbar \ov i} A_2 (r)}
\label{ulteq}
\end{eqnarray}
and setting $M_0 = 1$ and $M_j = 0 (j>1)$, we easily find the 
solution:
\begin{eqnarray}
& \tspino{F_1 (r)}{F_2 (r)}{D(r)} & = \tspino{\pm \lt( {\hbar \ov i} 
\ri) \lt[ 1 - {\lp \ov 2} \ri] {1 \ov 2 \sqrt{\lambda } r^2} - \lt( 
{\hbar \ov i} \ri)^2 \lt[ 1 - {\lp \ov 2} \ri] {N^2 \ov 4\lambda 
r^3}}{0}{0} \nonumber\\& & \mbox{} ~~~~ \cdot N^2 {c \ov r} e^{\pm {i 
\ov \hbar} 2 \sqrt{\lambda }r^* - i (l + 1) {\pi \ov 2} + O(\hbar )}.
\end{eqnarray}

Let us first consider the flat case $M=0$. By imposing the vanishing 
of the solutions, both ``even'' and ``odd'' for $r=R$, we obtain the spectrum:
\[
   2 {\sqrt{\lambda^0_{n,l}} \ov \hbar} R - \lp {\pi \ov 2} = n {\pi 
   \ov 2} 
\]
or 
\beq
\sqrt{\lambda^0_{n, l}} = {\hbar \ov 2R} ( n + l + 1 ) {\pi \ov 2}.
\label{p1}
\eeq
For $M\not=0$, one gets instead:
\beq
\sqrt{\lambda_{n, l}} = \sqrt{\lambda^0_{n, l}} {R \ov R^*} = 
\sqrt{\lambda^0_{n, l}} {1 \ov 1 + {2m \ov R} \ln \lt( {R \ov 2m} - 1 
\ri)},
\label{p2}
\eeq
which clearly shows
that, due to that presence of the wormhole, the 
gravitons' spectrum is redshifted by the quantity 
${R \ov R^*}={1 \ov 1 + {2m \ov R} \ln \lt( {R \ov 2m} - 1 
\ri)} $. In the next Section we shall analyse the consequences for the 
energy of our state of the results obtained so far.

\section{The energy of the quantized gravitational field around an 
ensemble of wormholes}

We have just seen that an external observer who looks into a finite 
spherical box (of radius $R$), centered around a wormhole, perceives 
that the gravitons contained in it are redshifted with respect to the 
gravitons in the absence of the wormhole. We wish now to 
determine the over all energy shift 
\beq
\Delta E(M) = E(M) - E(0)\nonumber
\eeq
between the gravitational quantum state in the box containing a 
wormhole and the one with no wormhole. In Section 2 we have seen that
\label{priqua}
\beq
E(M) = M + \sum_{n, l, m, \sigma} \sqrt{\lambda^\sigma_{n, l, m}}
\eeq
where $l,m$ are the angular quantum numbers, $n$ is the radial one
and $\sigma = \mp$ for ``even'' or ``odd'' solutions. Please note that 
we have come back to the natural units where $\hbar=c=1$. In flat 
space-time the energy is
\beq
E^0 = 
\displaystyle\sum_{n, l, m, \sigma} \sqrt{\lambda^{0 \sigma}_{n, l, 
m}}.
\eeq
Thus, from \reff{p1} and \reff{p2} we compute
\[
   \Delta E(M) = E(M) - E(0) \simeq M - {\pi \hbar \ov 2R^2} m {\hbox 
   {log}} {R \ov 2m} \sum_{n, l, m, \sigma} (n + l + 1).
\]
We must now sum over the quantum numbers, we get
\beq
\sum_{n, l, m, \sigma} (n + l + 1)=\sum_{nl}2(2l+1)(n+l+1).
\nonumber
\eeq
and introducing the momentum cut-off $\Lambda$\footnote{The problem of 
the cut-off, as already discussed, is a very important one. For 
consistency of our approach requires $G(\Lambda)\Lambda^2\le 1$, where
$G(\Lambda)$ is the value of the Newton constant at the cut-off 
$\Lambda$.} we have,
\beq
\Lambda={\pi\over4R}N\ge{\pi\over4R}(n+l+1).
\eeq
From:
\begin{eqnarray}
& & \sum_{n +l + 1 \leq N} 2(2l + 1) (l + n + 1) = \displaystyle{N^4 \ov 
2} - \displaystyle{17 \ov 18} N^3 + \displaystyle{N \ov 9} \nonumber\\& 
& \mbox{} ~~~~ \sim \displaystyle{N^4 \ov 2} = \displaystyle{64 R^4 \ov 
(\hbar \pi )^4} \Lambda^4.
\end{eqnarray}
we may finally write:
\beq
\Delta E (M) = M - {64 \Lambda^4 R^2 \ov \pi^3 \hbar^3} G M {\hbox 
{log}} {R \ov 2GM},
\label{de}
\eeq
or equivalently for the energy density:
\begin{equation}
{\Delta E(M)\over V}={M\over {4\pi\over3}R^3}-{36\Lambda^4\over\pi^4\hbar^3}
({GM\over R})\ln({R\over2GM}),
\label{newadd}
\end{equation}
which clearly shows that the ``stable'' gravitons' modes on the Schwarzschild 
background, due to the gravitational red-shift, have a zero-point 
energy smaller than in flat space-time, leading to the negative term 
in (\ref{de}). However we shall not analyse (\ref{de}) any further, 
for this expression is deficient in two important aspects: (i) it does not 
incorporate the (negative) classical contribution from the 
``unstable'' mode (\ref{unstable}); (ii) it only describes a single, isolated 
``wormhole'' which, in view of the large energy gains involved, if
certainly does not
correctly describe what we are after: a possible QG Ground State. In 
any event, the importance of (\ref{de}) lies in its exposing 
unequivocally the quantum mechanical instability of the classical 
Ground
State for the matterless universe: the (pseudo) euclidean space-time.

\section{The multi-wormhole state}

In order to have a realistic model for the Ground State of QG, as we 
have just argued, we must seek a more realistic classical solution of 
matterless Einstein's gravity. It is clear from what we have learned so 
far that the natural candidate is a ``condensed'' system of WH's, 
either 
gaseous or crystalline, of ADM mass $M$ and (average)
density $({1\over a})^3$, {\it a} being the interwormhole average distance. 
Even though no explicit solution, that we know of, has been given for 
the general case, one such classical state must certainly exist, for
by increasing $a$ one is led to the well-known Schwarzschild solution, and 
special configurations of the $n$-wormhole system have been studied 
and characterized in ref.\cite{gib}. From this study one confirms the 
intuitive expectation that the ADM energy of ``wormholes'' 
of mass $M$, for separations $a\gg 2GM$ (the Schwarzschild radius), coincides with 
the classical energy of a system of $n$ matter points interacting via 
the two-body Newtonian potential $U=-{GM^2\over a}$. Classically this 
configuration is unstable, since under the action of the Newtonian 
attraction the ``wormholes'' will not keep their (average) mutual 
distances $a$ but will tend to collapse to a single ``wormhole'' 
configuration, encompassing the given space-time 
$V$\footnote{Evidently part of the energy in the collapse must be 
radiated away, as the single wormhole mass in the volume $V$ is 
proportional to $n^{1\over3}M$ and not to $nM$.}. Quantum -mechanically 
the situation is different, for the Heisenberg principle  
puts in general bounds upon the interwormhole distance $a$. Indeed, for a 
two-wormhole system, from the Schr\"odinger equation
\beq
\lt[ - {\hbar^2 \ov 2\mu}\nabla^2_{\vec{d}} + U(d) \ri] u (\vec{d} ) = Eu 
(\vec{d}),
\eeq
where the reduced mass $\mu = {M \ov 2}$, we obtain for the lowest 
lying state:
\beq
E=-{GM^2\over2a_\circ},
\eeq
with the ``Bohr-radius''
\beq
a_\circ = {2\over GM^3}.
\label{aaaa}
\eeq
Thus, provided the ``Bohr radius'' $a_\circ$ exceeds the 
``coalescence'' distance $d_c=4GM$ (twice the Schwarzschild radius) the quantum 
mechanical system of $n$ wormholes is stable. This means that one must 
have:
\beq
a\ge a_\circ\ge 4GM,
\eeq
the maximum allowed density occurring when
\beq
a= a_\circ= 4GM,
\eeq
i.e.~for
\beq
M=\left({1\over2}\right)^{1\over4}G^{-{1\over2}}.
\label{yy}
\eeq
Extending (\ref{de}) to the $n$ ``wormholes'' case, and setting 
$R={a\over2}$, we obtain for the energy density difference $(a\ge 
4GM={2\over GM^3})$
\beq
{\Delta E\over V}\simeq {M\over a^3}\left[1-{GM\over4a}-{16\over\pi^3}
Ga^2\Lambda^4\ln\left({a\over4GM}\right)\right].
\label{eee}
\eeq
Thus, the energy density gain is maximum when $a$ is minimum, 
i.e.~when, according to (\ref{aaaa})
\beq
a=a_\circ = {2\over GM^3}.
\label{aaa}
\eeq
On the other hand, the quantum stability of the $n$-wormhole 
configuration
teaches us that the minimum value of the Bohr radius $a_\circ$ must 
equal $4GM$, implying that the maximum energy gain is achieved when 
the wormholes have, according to (\ref{yy}), an ADM-mass
$M=\left({1\over2}\right)^{1\over4}G^{-{1\over2}}=
\left({1\over2}\right)^{1\over4}m_p$, and an average distance
\beq
a_\circ = {2\over G\mu^3}=\left({1\over2}\right)^{7\over4}G^{-{1\over2}}
=\left({1\over2}\right)^{7\over4}l_p.
\eeq

We have just seen that, even without taking into account the 
(negative) contribution to the energy density of the unstable modes, a 
system of ``wormholes'' of ADM mass $M=2^{-{1\over4}}m_p$ 
($m_p=G^{-{1\over2}}\simeq 10^{19}$GeV, the Planck mass) at the average 
distance $\left({1\over2}\right)^{7\over4}l_p$ ($l_p=G^{1\over2}\simeq
10^{-33}$cm, the Planck length) realize, according to (\ref{eee}), a large 
energy (density) gain with respect to the Perturbative Ground State 
(PGS) of Quantum Gravity. 

We shall now try to answer the important 
question of the so far neglected contribution of the ``unstable'' mode
around a single wormhole. Let us consider around a wormhole of mass 
$M$, of Schwarzschild radius $2m=2GM$,
a spherical region of radius $R=\lambda m$ (${a\over m}>\lambda>2$).
The spherical region shell: $2m<r<\lambda m$ is the region (modulo 
the quantum fluctuations of the wormhole), which the asymptotic 
observer can probe, and in particular measure its (average) Riemann 
tensor.

The existence around the wormhole of an `` unstable'' solution of the 
gravitational field (which will obviously generalize to the case of 
many wormholes), found in section 3, guarantees us that such a mode 
will contribute a classical (negative) term to the ADM energy of the 
quantized gravitational field.
What kind and size of contribution? As argued in Section 3, without 
involving ourselves in the intractable expansion of the QG Hamiltonian to 3rd 
and higher orders, we may figure out the effect of the ``unstable'' 
mode in the following way. Let us call $\phi^{(u)}_{ij}$ the 
normalized unstable mode, which according to the developments in section 4, can be 
expressed in terms of the function $H(r)$ given (\ref{nellaprima}) and 
the constraint
function $U(r)$, appearing in (\ref{unstable}). Its contribution $h_{ij}(x)$ to 
the quantized gravitational field is thus 
\beq
h^u_{ij}=A\phi^u_{ij},  
\label{999}
\eeq
where $A$ is a real constant whose value shall be determined in such a 
way as to minimize the Riemann tensor associated with the metric field
\beq
g_{ij}=\eta^s_{ij}+h^u_{ij}.
\eeq
Thus we examine the metric:
\[
g_{{11}}=\left (1-{\frac {2m}{r}} \right)^{-1} -2\,A{
\frac {1}{\sqrt {1-{\frac {2m}{r}}}}}{r}^{-3}
\]

\[
g_{{22}}={r}^{2}+A\sqrt {1-{\frac {2m}{r}}}{r}^{-1}
\]

\[
g_{{33}} = \left( {r}^{2}+A\sqrt {1-{\frac {2m}{r}}}{r
}^{-1}\right) \left( \sin \theta \right)^{2}
\]
which yields the non-vanishing components of the 3-dimensional Riemann 
tensor 
are:
\begin{eqnarray}
R_{121}^{~~~2} = & & \left(123\,{m}^{2}{A}^{2}{r}^{3}+12\,{m}^{2
}{A}^{3}\sqrt {-{\frac {-r+2\,m}{r}}}-116\,
m{A}^{2}{r}^{4}-4\,m{r}^{10} \ri. \nonumber\\& & \mbox{} \lt. 
+14\,{r}^{7} \sqrt {-{\frac {-r+2\,m}{r}}}mA-8\,m{A}^{3}
r\sqrt {-{\frac {-r+2\,m}{r}}} + 
27\,{A}^{2}{r}^{5}\right) \nonumber\\& & \mbox{} ~~~ 
\left(4\,{r}^{4}\sqrt {-{\frac {-r + 2m}{r}}} 
8\,Ar+16\,Am\right )^{-1} {\frac {1}{\sqrt {-{\frac {-r+2\,m}{r}}}
}} \nonumber\\& & \mbox{} ~~~ \left ({r}^{3}+A\sqrt {-{\frac 
{-r+2\,m}{r}}}\right )^{-2}{r}^{-3}
\end{eqnarray}

\begin{eqnarray}
R_{232}^{~~~3} = & & \left(-12\,{r}^{5}\sqrt {-{\frac {-r+2\,m}{
r}}}Am-9\,{A}^{2}{r}^{3}+40\,{A}^{2}{r}^{2}
m-53\,{A}^{2}{m}^{2}r \ri. \nonumber\\& & \mbox{} ~~~ \lt. + 8\,m{r}^{8} 
- 16\,{m}^{2}{r}^{7}+24\,A{m}^{2}{r}^{4}\sqrt {-{
\frac {-r+2\,m}{r}}}+18\,{A}^{2}{m}^{3}
\right)1/4{r}^{-2} \nonumber\\& & \mbox{} ~~~ \left( {r}^{3}+A\sqrt {-{
\frac {-r+2\,m}{r}}}\right )^{-1}{\frac {1}
{\sqrt {-{\frac {-r+2\,m}{r}}}}} \nonumber\\& & \mbox{} ~~~ \left ({r}^
{4}\sqrt {-{\frac {-r+2\,m}{r}}}-2\,Ar+4\,A
m\right )^{-1}
\end{eqnarray}

Averaging these expression in a spherical shell $2m<r<3m$, we have 
checked numerically that for $A\simeq 14(GM)^3$,
\beq
\left|{\langle R^2_{121}\rangle\over \langle 
R^{2(s)}_{121}\rangle}\right|\simeq
\left|{\langle R^3_{232}\rangle\over \langle 
R^{3(s)}_{232}\rangle}\right|\ll 1,
\eeq
clearly showing the possibility that in a shell of radius $O(m)$ around 
a ``wormhole'' the ``unstable'' mode screens completely the classical 
curvature induced by the ``wormhole'' itself. In view of the general 
result that the zero-curvature metric realizes the minimum energy 
density of the classical gravitational field, the maximum screening of 
the curvature existing around of a ``wormhole'' allows the 
``unstable'' mode of the quantized gravitational field to produce the 
minimization of the ADM (classical) energy that characterizes the 
Ground State of Quantum Gravity.

\section{The gas of wormholes: a possible ground state of quantum 
gravity}

Let us now, in this concluding Section, try to distill the main points 
and results of this work. In order to have a better understanding of 
what has been achieved, let us first describe again the strategy we 
have adopted to probe the stability of the classical ground state --
the flat Minkowskian background -- subject to the fluctuations of the 
quantized gravitational field.

For definiteness' sake, for the classical solution of matterless 
gravity, our Ansatz for the classical background of QG, we have chosen 
the Schwarzschild solution but other possible 
starting points could be the Reissner-Nordstrom or the Kerr solution. 
From the classical standpoint such a background cannot be a good model 
for the vacuum, for it is well-known that the ADM\cite{a} mass of
such solution is positive ($E_{ADM}=M$), whereas $E_{ADM}=0$ for flat
space-time. But for the quantized gravitational field things may in 
principle be quite different, for $E_{ADM}$ receives contributions not 
only from the {\it classical} background field but also from the 
quantum fluctuations described by the quantum field $h_{ij}$ related 
to the metric field by:
\beq
g_{ij} = \eta^{(s)}_{ij} + h_{ij},
\label{gh}
\eeq
$\eta^{(s)}_{ij}$ being the Schwarzschild solution, fully characterized by the 
ADM-mass $M$ (and Schwarzschild radius $r_s=2GM=2m$). And it may happen, as it 
has been discovered in ref.\cite{gp}, that the full energy $E(M)$ 
classical plus quantum, of the configuration (\ref{gh}) may turn out 
to be smaller than quantum energy $E(0)$ of the Perturbative Ground 
State (PGS), where $\eta^s_{ij}\rightarrow \delta_{ij}$. If this 
indeed happened one would have finally got rid of the embarrassing PGS 
and of the unrenormalizable perturbative field theory that inevitably
obtains upon it.

As stressed in the Introduction, our hope that such strategy might 
lead to a similar situation, thus giving us a sensible QG, was 
based on a similar analysis performed on a non-abelian gauge theory, 
the $SU(2)$ pure Yang-Mills theory (akin to QCD)\cite{gp}, where it 
was discovered that the energy $E(B)$, of the quantized field 
fluctuating around a classical solution, describing a constant chromomagnetic
field $B$, was in fact lower than the PGS-energy $E(0)$. In the course 
of this analysis it was also discovered that around this non-trivial 
classical background there exist ``unstable'' modes (whose frequencies 
of small oscillations are imaginary) that profoundly modify the 
structure of the theory even at very high frequencies, fatally 
undermining the paradigmatic Asymptotic Freedom\cite{free}. Of course, 
our hope was strengthened by the observation that QG may be looked at 
as a non-abelian gauge theory whose group is the Poincar\'e group, 
acting upon the tangent Minkowskian spaces.

The analysis reported in Sections 4 and 5 reveals that around an 
isolated ``wormhole'':
\begin{itemize}
\begin{enumerate}

\item there exist an S-wave ``unstable'' mode [see (\ref{nellaprima}), 
(\ref{unstable}) and (\ref{e})];

\item the high energy ``stable'' modes are red-shifted with respect to 
the gravitons of flat Euclidean space, and realize the large energy gain 
$\Delta E(M)$ of eq.(\ref{de}).

\end{enumerate}
\end{itemize}

As a result the PGS is clearly seen to be unstable: the quantized 
field finds it energetically advantageous to ``concentrate'' in a 
``wormhole'' of radius $2GM$ and ``mass'' $M$, which redshifts the 
zero-point modes of the gravitational field, lowering in this way 
their energy density.

But, as shown in Section 6, the single ``wormhole'' configuration 
certainly does not minimize the energy of the gravitational field, for 
the appearance of several other ``wormholes'' of mass $M$ at an 
appropriate inter-wormholes (average) distance $a$ - a gas or a lattice 
of ``wormholes''-- leads to a decrease of the energy density [see 
eq.(\ref{eee})] provided, however, that $a$ is bigger than the distance 
$d=4GM$, at which two of them coalesce into a single ``wormhole''. 
Taking into account such constraint, from (\ref{eee}) and (\ref{aaa}) 
we derive 
that an ensemble of ``wormholes'' with mass $M\sim G^{-{1\over2}}=m_p$ 
at the (average) distance $a\simeq G^{1\over2}=l_p$ minimizes the part 
of the energy density that takes no account of the (negative) 
classical energy density stemming from the so far neglected 
``unstable'' modes.

At the end of section 6 by a numerical analysis we show that the 
variational amplitude $A$ of the ``unstable'' mode [see eq.(\ref{999})] can 
be chosen in such a way that the {\it classical} part of the Riemann 
tensor, averaged over a spherical shell $(2m<r<3m)$ around a 
``wormhole'' is much smaller than the averaged Riemann tensor of the 
Schwarzschild solution. In this way the energy density can be further lowered, 
obtaining a state of the quantized gravitational field, fluctuating 
around a gas of wormholes of Planck mass with Planck distance 
separation, whose energy density is way below the energy density of 
the PGS: an excellent candidate for the GS of QG. 

Why do we believe that this latter statement is of significance for 
our understanding of the fundamental laws of physics? First of all we 
must note, as emphasized in the Introduction, that the gas of 
``wormholes'' that minimizes the energy density of the quantized 
gravitational field realizes the remarkable intuition of J.A.~Wheeler 
that at the Planck distance continuous space may dissolves in a kind of 
{\it foam} whose voids have the Planck size $l_p$. Even though 
Wheeler's {\it foam} can only be seen when our space resolution 
reaches $l_p$, just an impossible dream for presently available 
``microscopes'' (the high energy accelerators), its effect on the 
structure and self-consistency of the QFT's of the Standard Model 
and for QG itself is enormous. In fact the space-structure upon which 
the observable quantum fields are defined turns out to be essentially 
discrete, comprising the Planck size space domains located at the 
interstices of the Planck size WH's which form the ``points'' of a 
``Planck Lattice'' (PL). 

This latter statement might appear somewhat 
surprising, in view of the fact that the idea of a "foam" seems to retain
a basic element of continuity of space-time which is lost in a lattice
formulation. However, if we now look at the physics of the phenomenon of 
"WH-condensation" which, we have shown, leads possibly to the real GS of QG,
we realize at once that no meaning can be attached to the value of the 
gravitational field or, for that matter, of any other physical field, 
in space-time domains whose (linear) size is less than the Planck length $l_p$.
And this for two reasons, first domains of such size are hidden beyond 
the horizen of the single WH and cannot affect the physics of the observable
space-time domains (of the same size) apart from their gravitational effects,
that are however compeltely screened by the unstable modes; and second the
gas-like structure of the system of WH's with their fast quantum fluctuations
cannot but ``average'' any ``local'' field over Planck-sized space-time 
domains, thus rendering unobservable, indeed unphysical, all field modes 
whose wave-lengths are smaller than $l_p$. Without such modes in the spectrum 
of any physical field it obviously follows that there is no way to probe 
space-time deeper than $l_p$, thus allowing us to effective describe 
space-time as  an essentially discrete structure, a random lattice of average 
lattice constant $a=l_p$.  

Even without going into the details of the 
significant advances in a fundamental understanding of the SM that a 
research program based on the PL achieves\footnote{A concise 
bibliography is to be found in ref.\cite{xuepre}}, we can realize at 
once that the discretization of space in a PL introduces a fundamental
momentum cut-off $\Lambda\simeq m_p$, the Planck mass, thus removing 
at a basic level all divergences of QFT's, which since their 
formulations at the end of the 20's, have been a devastating 
conceptual difficulty, that the ``conventionalistic'' way out of 
``renormalization'' could only throw ``under the rug'', arousing the 
anger of great minds such as Dirac's. With a PL, a faithful and 
adequate representation of the structure of the non-perturbative GS of 
QG, the ghosts of renormalization theory, such as the Landau's ghost, 
disappear beyond the horizon of the WH's, leaving us a decent 
perturbative expansion for the QFT's of the Standard Model, whose 
logarithmic divergences
get converted to small corrections once one sets 
$\Lambda=m_p$\footnote{This obviously does not happen for scalar 
theories, such as the Higgs model, where there appear ``quadratic'' 
divergences
in $\Lambda^2$,which retain their devastating character, thus exposing
their unplausibility.}. But this is not all, also the observable part 
of the quantized gravitational field, living on the PL, is cut -off at 
$m_p$, and the ``nonrenormalizable'' divergences $O[(G\Lambda^2)^n]$ 
become small corrections once the providential
powers of $2\pi$ in the denominators, stemming from momentum 
integrations, are fully taken into account\footnote{As we have been
emphasizing in this work, the 
cut-off at $m_p$ fully justifies our neglect in the Hamiltonian of 
terms of order higher than two.}. In this way, QG leaves the 
``limbo'' of the theories of uncertain ``self-consistency'' to assume the 
status of a perfectly well defined QFT, with a well-defined 
perturbative S-matrix, whose quantum effects are governed in a 
well-defined fashion by a very small coupling constant $G$, the Newton 
constant. And in view of all this one needs only the ``Ockham's 
razor'' to cut away much of the sophisticated theoretical 
developments, based on ``superstrings'', of the last 15 years. 

Let us 
end this paper by asking ourselves whether there remain open problems.
Let us emphasize that we deem our results qualitatively robust, 
especially as for as the derivation of a quantum foam is concerned; an 
interesting open problem is then certainly the extension of the analysis to 
more complicated vacuum solutions, such as the Reissner-Nordstrom and 
Kerr metric. A more refined analysis, aimed at determining the 
detailed structure of the quantum foam and its approximation as a PL, 
is certainly needed. We hope to come back to this problem in the near 
future. Finally we think that ``vexata quaestio'' of the ``cosmological
constant'' $\Lambda_c$ may have an answer within the framework of this 
paper, and that the ``unstable'' quantized modes of the gravitational 
field around each WH may have a lot to do with the surprising, and 
extremely welcome, negligible value that $\Lambda_c$ has in our Universe.

We acknowledge the collaboration of S.~Cacciatori and I.~Spagnolatti 
in the early stage of this work.




\section{Appendix A: Christoffel Symbol and Ricci tensor for the background 
field}

We give detail calculations of some fundamental quantities generated 
by ${\bbox {\eta}}$. We call 
\begin{eqnarray}
\bbox{\eta} = \left (\matrix {& {1 \ov 1 - {2m \ov r}} & 0 & 0 \cr & 0 & 
r^2  & 0  \cr & 0 & 0 & r^2 \sin^2 \theta }\right) .
\end{eqnarray}
For more convenience, we select the cartesiane coordinate:
$ x^1 = r\sin \theta  \cos \varphi, x^2 = r\sin \theta  
\sin \varphi, x^3 = r \cos \theta $, and
\[ 
d \xv = \lt( {x^1 \ov r} dr + x^3 \cos \varphi d\theta  - x^2 
d\varphi ; {x^2 \ov r} dr + x^3 \sin \varphi d\theta  + x^1 d\theta ; 
{x^3\ov r} dr - r\sin \theta  d\theta  \ri) 
\]
and then
\beq
ds^2 = \lt\{ \lt( 1 - {2m \ov r} \ri)^{-1} - 1 \ri\} {(\xv \cdot d \xv 
)^2 \ov r^2}- (d \xv )^2
\eeq
where 
\[ 
   \xv \cdot d\xv = \sum_a x^a dx^a. 
\]
As a result,
\[ 
\eta_{ab} = \delta_{ab} + {x_a x_b \ov r^2} \lt( {1 \ov 1 - {2m \ov 
r}} -1 \ri) 
\] 
and, in particular, we obtain 
\beq
x_r = g_{rs} x^s = x^r \lt( 1 - {2m \ov r} \ri) .
\eeq
We first calculate the Christoffel symbol,
\beq
\Gamma^{r(s)}_{ab} = {1 \ov 2} \eta^{rs} \Big( \de_a \eta_{sb} + \de_b 
\eta_{sa} - \de_s \eta_{ab} \Big).
\eeq
Since
\beq
\de_a \eta_{sb} = {2m \ov r^3 \lt( 1 - {2m \ov r} \ri)} \Big( 
\delta_{as} x_b + \delta_{ab} x_s \Big) - {x_b x_s 2m \ov r^b \lt( 1 - 
{2m \ov r } \ri)^2 } \Big( 3r^2 - 4mr \Big) {x_a \ov r}, 
\eeq
and obtain,
\begin{eqnarray}
\Gamma^{r~(s)}_{ab} & = & {2m \ov r^3} {\delta_{ab} x^r \ov \lt(1 - 
{2m \ov r} \ri) } \lt[ {3 \ov 1 - {2m \ov r} } - {4m \ov r \lt( {2m \ov r} 
\ri) } \ri] \nonumber\\ & & \mbox{} = {2m \ov r^3} {\delta_{ab} x^a \ov 1 
- {2m \ov r} } - {m x_a x_b x^r \ov r^5 \lt( 1 - {2m \ov r} \ri) } 
\lt[ 3 + {2m \ov r \lt( 1 - {2m \ov r} \ri) } \ri] . 
\end{eqnarray}
In the coordinate $(r, \theta  , \varphi )$ we have:
\begin{eqnarray}
\Gamma^1_{lm} = \left (\matrix {& {-m \ov r(r - 2m )} & 0 & 0 \cr & 0 & 
- (r-2m)  & 0  \cr & 0 & 0 & -(r - 2m) \sin^2 \theta  } \right), 
\end{eqnarray}
\begin{eqnarray}
\Gamma^2_{lm}  = \left (\matrix {& 0 & {1 \ov r} & 0 \cr & {1 \ov r} & 
0  & 0  \cr & 0 & 0 & - \sin \theta  \cos \theta  }\right), 
\end{eqnarray}
\begin{eqnarray}
\Gamma^3_{lm} = \left (\matrix {& 0 & 0 & {1 \ov r} \cr & 0 & 
0 & {\hbox{cotg}} \theta   \cr & {1 \ov r} &  {\hbox{cotg}} \theta  & 
0 }\right).
\end{eqnarray}
We can then calculate the Ricci tensor:
\beq
R^{(s)}_{ab} = \Gamma^{r~(s)}_{ra,b} - \Gamma^{r~(s)}_{ab,r} + 
\Gamma^{r~(s)}_{sa} \Gamma^{s~(s)}_{rb} - \Gamma^{r~(s)}_{ab} 
\Gamma^{r~(s)}_{rs}.
\eeq
Since
\beq
\Gamma^{r~(s)}_{ra} = - {m \ov r^3} {x_a \ov 1 - {2m \ov r}},
\eeq
we find:
\beq
\Gamma^r_{ra,b} = - {m\delta_{ab} \ov r^3 \lt( 1 - {2m \ov r} \ri) } 
+ {m x_a x_b \ov r^5 \lt( 1 - {2m \ov r} \ri)^2 } \lt( 3 {4m \ov r} 
\ri).
\eeq
and
\begin{eqnarray}
\Gamma^{r~(s)}_{ab,s} & = & {2m \ov r^3} \delta_{ab} \delta_{rs} - {6m 
\ov r^5} \delta_{ab} x_r x_s - {m \ov r^5} \lt[ 3 + {2m \ov r \lt( 1 - 
{2m \ov r} \ri) } \ri] \Big( \delta_{as} x_b x_r \nonumber\\& & \mbox{} 
+ \delta_{sb} x_a x_r + \delta_{sr} x_a x_b \Big) + {m x_a x_b x_r x_s 
\ov r^7} \lt[ 15 + {6r^5 - 10mr^4 \ov r^5 \lt( 1 - {2m \ov r} \ri)^2 
}\ri]. 
\end{eqnarray}
From this equation, in particular we obtain:
\begin{eqnarray}
\Gamma^{r~(s)}_{ab,r} & = & - 5x_a x_b {m \ov r^5} \lt[ 3 + {2m \ov r 
\lt( 1 - {2m \ov r} \ri) } \ri] + {m x_a x_b \ov r^5} \lt[ 15 + {6r - 
10m \ov r \lt( 1 - {2m \ov r} \ri)^2 } \ri] \nonumber\\& & \mbox{} = 
{2m^2 x_a x_b \ov r^6 \lt( 1 - {2m \ov r} \ri)^2 }, 
\end{eqnarray}
and
\begin{eqnarray}
\Gamma^{r~(s)}_{sa} \Gamma^{s~(s)}_{rb} & = & \lt\{ {2m \ov r^3} 
\delta_{sa} x_r - {m x_a x_s x_r \ov r^5} \lt[ 3 + {2m \ov r \lt( 1 - 
{2m \ov r }\ri) } \ri] \ri\} \nonumber\\& & \mbox{} \cdot \lt\{ {2m \ov r^3} 
\delta_{rb} x_b - {m x_r x_b x_s \ov r^5} \lt[ 3 + {2m \ov r \lt( 1 - 
{2m \ov r }\ri) } \ri] \ri\} \nonumber\\& & \mbox{} = x_a x_b {m^2 \ov 
r^6} \lt[ 1 + {2m \ov r \lt( 1 - {2m \ov r} \ri) } \ri]^2 
\nonumber\\& & \mbox{}= x_a x_b {m^2 \ov r^6} {1 \ov \lt( 1 - {2m \ov r} 
\ri)^2 }. 
\end{eqnarray}
At the end, we have
\begin{eqnarray}
-\Gamma^{a~(s)}_{ab} \Gamma^{s~(s)}_{rs} & = & \lt\{ {2m \ov r} 
\delta_{ag} x_r - {m x_a x_b x_r \ov r^5} \lt[ 3 + {2m \ov r \lt( 1 - 
{2m \ov r} \ri) } \ri] \ri\} \nonumber\\& & \mbox{} ~~~~ \cdot {m \ov 
r^3} {x_a \ov 1 - {2m \ov r}} = {2m^2 \ov r^4} {\delta_{ab} \ov \lt( 1 
- {2m \ov r} \ri) } \nonumber\\& & \mbox{} - {m^2 \ov r^6} {x_a x_b \ov 
\lt( 1 - {2m \ov r} \ri) } \lt[ 3 + {2m \ov r \lt( 1 - {2m \ov r} \ri) } 
\ri], 
\end{eqnarray}
and as a result:
\begin{eqnarray}
R^{(s)}_{ab} & = & - {m \delta_{ab} \ov r^3 \lt( 1 - {2m \ov r} \ri) } 
- {4m^2 x_a x_b \ov r^6 \lt( 1 - {2m \ov r} \ri)^2 } + {3m x_a x_b \ov 
r^5 \lt( 1 - {2m \ov r} \ri)^2} \nonumber\\& & \mbox{} - {2m^2 x_a x_b 
\ov r^6 \lt( 1 - {2m \ov r} \ri)^2 } + {x_a x_b m^2 \ov r^6 \lt( 1 - 
{2m \ov r} \ri)^2 } + {2m^2 \ov r^4} {\delta_{ab} \ov 1 - {2m \ov 
r}} \nonumber\\& & \mbox{} - {m^2 \ov r^6} {x_a x_b \ov 1 - {2m \ov r} 
} \lt[ 2 + {1 \ov \lt( 1 - {2m \ov r} \ri) } \ri] = - {m\ov r^3} 
\delta_{ab} - {2m^2 \ov r^6} {x_a x_b \ov 1 - {2m \ov r} } \nonumber\\& 
& \mbox{} + {3m x_a x_b \ov r^5 \lt( 1 - {2m \ov r} \ri)^2 } = - {m 
\ov r^3} \eta_{ab} + {3m x_a x_b \ov r^5 \lt( 1 - {2m \ov r} \ri) }. 
\end{eqnarray}
Owning to
\[ {x_a x_b \ov r^2} {2m \ov 1 - {2m \ov r}} = \eta_{ab} - 
\delta_{ab}, \]
the Ricci tensor of the background field is:
\[ 
R^{(s)}_{ab} = {m \ov r^3} \Big( - \eta_{ab} + {3r \ov 2m} 
\eta_{ab} - {3r \ov 2m} \delta_{ab} \Big).
\]
Going back to the coordinate $(r, \theta  , \varphi )$ in the matrix 
form, we have:
\begin{eqnarray}
R^{a~(s)}_b = \left (\matrix {& -{2m \ov r^2} & 0 & 0 \cr & 0 & 
{m \ov r^2} & 0 \cr & 0 &  0 & {m \ov r^2} }\right) .
\label{ultaa}
\end{eqnarray}
In particular we find that the scalar of curvature is zero: $R^{(s)} = 0$.

\section{Appendix B: Decomposing the  Regge-Wheeler matrix}

Using the technique due to T.Regge and 
J.A. Wheeler \cite{rw}, we separate the angular part from radial part
in the 3-dimension tensors $\phi^i_j (r, \theta  , \varphi )$,
vector $v_i (r, \theta,\varphi )$ and scalar $\tau (r, \theta  , \varphi )$.

\noindent{\it The even part:}
\beq
\phi^1_1 (r, \theta  , \varphi ) = H(r) Y_{lm} ( \theta  , \varphi ),
\eeq
\beq
\phi^2_2 (r, \theta  , \varphi ) = \Big( G_1 (r) + G_2 (r) \de^2_\theta  
\Big) Y_{lm} (\theta  , \varphi ),
\eeq
\beq
\phi^3_3 (r, \theta  , \varphi ) = \lt( G_1 (r) + {G_2 (r) \ov \sin^2 
\theta  }\de^2_\varphi + G_2 (r) \cot \theta  \de_\theta  \ri) Y_{lm} 
(\theta  , \varphi ),
\eeq
\beq
\phi^2_1 (r, \theta  , \varphi ) = K(r) \de_\theta  Y_{lm} (\theta  , \varphi 
),
\eeq
\beq
\phi^3_1 (r, \theta  , \varphi ) = {K(r) \ov \sin^2 \theta  } 
\de_\varphi Y_{lm} (\theta  , \varphi ),
\eeq
\beq
\phi^3_2 (r, \theta  , \varphi ) = {G_2 (r) \ov \sin^2 \theta  } \Big( 
\de_\theta  - \cot \theta  \Big) \de_\varphi Y_{lm} (\theta  , \varphi 
),
\eeq
\beq
v_1 (r, \theta  , \varphi ) = U(r) Y_{lm} (\theta  , \varphi ),
\eeq
\beq
v_2 (r, \theta  , \varphi ) = V(r) \de_\theta  Y_{lm} (\theta  , \varphi 
),
\eeq
\beq
v_3 (r, \theta  , \varphi ) = V(r) \de_\varphi Y_{lm} (\theta  , \varphi 
),
\eeq
\beq
\tau (r, \theta  , \varphi ) = T(r) Y_{lm} (\theta  , \varphi ).
\eeq
{\it The odd part:}
\beq
\phi^1_1 (r, \theta  , \varphi ) = 0,
\eeq
\beq
\phi^2_2 (r, \theta  , \varphi ) = F_2 (r) {1 \ov \sin \theta  } \Big( 
\de_\theta  - \cot \theta  \Big) \de_\varphi Y_{lm} (\theta  , \varphi 
),
\eeq
\beq
\phi^3_3 (r, \theta  , \varphi ) = - \phi^2_2 (r, \theta  , \varphi ),
\eeq
\beq
\phi^2_1 (r, \theta  , \varphi ) = - {F_1 (r) \ov \sin \theta  } 
\de_\varphi Y_{lm} (\theta  , \varphi ),
\eeq
\beq
\phi^3_1 (r, \theta  , \varphi ) = {F_1 (r) \ov \sin \theta  } 
\de_\theta Y_{lm} (\theta  , \varphi ),
\eeq
\beq
\phi^3_2 (r, \theta  , \varphi ) = {1 \ov 2} {F_2 (r) \ov \sin \theta  } 
\lt( {1 \ov \sin^2 \theta  } \de^2_\varphi + \cot \theta  \de_\theta  - 
\de^2_\theta  \ri) Y_{lm} (\theta  , \varphi ),
\eeq
\beq
v_1 (r, \theta  , \varphi ) = 0,
\eeq
\beq
v_2 (r, \theta  , \varphi ) = D(r) \lt( -{1 \ov \sin \theta  } \ri) 
\de_\theta  Y_{lm} (\theta  , \varphi ),
\eeq
\beq
v_3 (r, \theta  , \varphi ) = D(r) \sin \theta  \de_\theta  Y_{lm} (\theta  , 
\varphi ).
\eeq
Obviously, the scalar has no odd part.

\section{Appendix C: Explicit form of the eigenequations of Hamiltonia}

We give the great details of calculations for obtaining the equations 
of Hamiltonia eigenfunctions and constrains.

\noindent{\it Laplacians of tensors}

We show how to operate the Laplaciano $\De^2 
\equiv \eta^{ab} \De_a \De_b$ where the derivatives are taken 
covariantly with respect to the matrix $\eta_{ab}$.

For a tensor with rank 2, we find:
\begin{eqnarray}
& \nabla^2 h^i_j & = \eta^{ab} h^i_{j \mid ab} =\eta^{ab} \left[ h^i_{j 
\mid a,b} +\Gamma^i _{rb} h^r_{j \mid a} -\Gamma^r _{jb} h^i_{r \mid 
a} -\Gamma^r _{ab} h^i_{j \mid r} \right]\nonumber\\& & \mbox{} = 
\eta^{ab} \left[ 
h^i_{j,ab} +\left( h^r_j \Gamma^i_{ra} -h^i_r \Gamma^r_{ja} 
\right)_{,b} +\ga{i}{r}{b} h^r_{j,a} + \ga{i}{r}{b} \ga{r}{s}{a} h^s_j 
-\ga{i}{r}{b} \ga{s}{j}{a} h^r_s \right. \nonumber\\& & \mbox{} -  \left. 
\ga{r}{j}{b} \ga{i}{s}{a} h^s_r 
-\ga{r}{j}{b} h^i_{r,a} +\ga{r}{j}{b} \ga{s}{r}{a} h^i_s -\ga{r}{a}{b} 
h^i_{j,r} -\ga{r}{a}{b} \ga{i}{s}{r} h^s_j + \ga{r}{a}{b} 
\ga{s}{j}{r} 
h^i_s \right] \nonumber\\& & \mbox{} = \eta^{ab} \left[ h^i_{j,ab} 
+ 2 \ga{i}{r}{a} h^r_{j,b} -2\ga{r}{j}{a} h^i_{r,b}  +h^r_j {\ga{i}{r}{a}}_{,b} 
-h^i_r {\ga{r}{j}{a}}_{,b} +\ga{i}{r}{b} \ga{r}{s}{a} h^s_j \right. 
\nonumber\\& & \mbox{} - \left. 2\ga{i}{r}{b} \ga{s}{j}{a} h^r_s + \ga{r}{j}{b} 
\ga{s}{r}{a} h^i_s -\ga{r}{a}{b} h^i_{j,r} -\ga{r}{a}{b} \ga{i}{s}{r} h^s_j 
+\ga{r}{a}{b} \ga{s}{j}{r} h^i_s \right].
\label{dedhij}
\end{eqnarray}
For a covariant vector, we find:
\begin{eqnarray}
& \nabla^2 B_a & = \eta^{bc} B_{a \mid bc} =\eta^{bc} \left( B_{a 
\mid b,c} -\ga{r}{a}{c} B_{r \mid b} -\ga{r}{b}{c} B_{a \mid r} 
\right) \nonumber\\& & \mbox{} = \eta^{bc} \left[ \left( B_{a,b} 
-\ga{s}{a}{b} B_s \right)_{,c} -\ga{r}{a}{c} \left( B_{r,b} 
-\ga{s}{r}{b} B_s \right) -\ga{r}{b}{c} \left( B_{a,r} -\ga{s}{a}{r} B_s 
\right) \right] \nonumber\\& & \mbox{} = \eta^{bc} \left[ B_{a,bc} 
-\ga{s}{a}{b} B_{s,c} -{\ga{s}{a}{b}}_{,c} B_s 
-\ga{r}{a}{c} B_{r,b} +\ga{r}{a}{c} \ga{s}{r}{b} B_s -\ga{r}{b}{c} B_{a,r} 
\right. \nonumber\\& & \mbox{} + \left. \ga{r}{b}{c}\ga{s}{a}{r} B_s 
\right].
\end{eqnarray}
Using, in the place of $\Gamma^a_{bc}$, the vectors calculated 
in Appendix A, We easily find:
\begin{eqnarray}
\De^2 \phi^1_1 & = & \lt[ {r - 2m \ov r} \de^2_r + {1 \ov r^2} 
\de^2_\theta  + {1 \ov r^2 \sin^2 \theta  } \de^2_\varphi + {2r - 3m \ov 
r^2} \de_r \ri. \nonumber\\& & \mbox{} \lt. + \cot \theta  \de_\theta - 
4 {r - 2m \ov r^3} \ri] \phi^1_1 - {4(r - 2m) \ov r^2} \Big( \de_\theta 
+ \cot \theta  \Big) \phi^2_1 \nonumber\\& & \mbox{} - {4(r - 2m)\ov 
r^2} \phi^3_1 + 2 {r - 2m \ov r^3} \phi^2_2 + 2 {r - 2m \ov r^3} 
\phi^3_3;
\label{dedphi}
\end{eqnarray}
\begin{eqnarray}
\De^2 \phi^2_2 & = & \lt[ {r - 2m \ov r} \de^2_r + {1 \ov r^2} 
\de^2_\theta  + {1 \ov r^2 \sin^2 \theta  } \de^2_\varphi + {2r - 3m \ov 
r^2} \de_r + \cot \theta  \de_\theta \ri. \nonumber\\& & \mbox{} \lt. - 
2 {r - 2m \ov r^3} - {2\cot^2 \theta  \ov r^2} \ri] \phi^2_2 + {4(r - 
2m) \ov r^2} \de_\theta  \phi^2_1 \nonumber\\& & \mbox{} - {4 
\ov r^2} \cot \theta  \de_\varphi \phi^3_2 + 2 {r - 2m \ov 
r^3} \phi^1_1 + 2 {\cot^2 \theta  \ov r^2} \phi^3_3;
\end{eqnarray}
\begin{eqnarray}
\De^2 \phi^3_3 & = & \lt[ {r - 2m \ov r} \de^2_r + {1 \ov r^2} 
\de^2_\theta  + {1 \ov r^2 \sin^2 \theta  } \de^2_\varphi + {2r - 3m \ov 
r^2} \de_r + \cot \theta \de_\theta \ri. \nonumber\\& & \mbox{} \lt. - 
2 {r - 2m \ov r^3} - {2\cot^2 \theta  \ov r^2} \ri] \phi^3_3 
+ {4(r - 2m) \ov r^2} \de_\varphi \phi^3_1 \nonumber\\& & \mbox{} + 
{4\cot \theta  \ov r^2} \de_\varphi \phi^3_2 + 4\cot \theta  {r - 2m \ov 
r^3} \phi^2_1 \nonumber\\& & \mbox{} + 2 {r - 2m \ov r^3} \phi^1_1 + 2 
{cotg^2 \theta  \ov r^2} \phi^2_2;
\end{eqnarray}
\begin{eqnarray}
& & \De^2 \phi^2_1 = \lt[ {r - 2m \ov r} \de^2_\varphi + {1 \ov r^2} 
\de^2_\theta  + {1 \ov r^2 \sin^2 \theta  } \de^2_\varphi + {4r - 5m \ov 
r^2} \de_r \ri. \nonumber\\& & \mbox{} ~~~~ ~~ \lt. + {\cot \theta  \ov 
r^2} \de_\theta  - {3r - 9m \ov r^3} - {\cot^2 \theta  \ov r^2} \ri] 
\phi^2_1 - {2 \ov r^3} \de_\theta  \phi^1_1 \nonumber\\& & \mbox{} ~~~~ ~~ 
- 2{\cot \theta  
\ov r^2} \de_\varphi \phi^3_1 - {2 \ov r^3} \de_\varphi \phi^3_2 - {2 
\ov r^3} \Big( \cot \theta  + \de_\theta \Big) \phi^2_2 + {2 \cot 
\theta  \ov r^3} \phi^3_3;
\end{eqnarray}
\begin{eqnarray}
\De^2 \phi^3_1 & = & \lt[ {r - 2m \ov r} \de^2_r + {1 \ov r^2} 
\de^2_\theta  + {1 \ov r^2 \sin^2 \theta  } \de^2_\varphi + {4r - 5m \ov 
r^2} \de_r + {3\cot \theta  \ov r^2} \de_\theta \ri. \nonumber\\& & \mbox{} 
\lt. - 4 {r - 9m \ov r^3} \ri] \phi^3_1 + {2 \ov r^3 \sin^2 \theta } 
\de_\varphi \phi^1_1 + {2\cot \theta  \ov r^2 \sin^2 \theta  } 
\de_\varphi \phi^2_1 \nonumber\\& & \mbox{} 
- {2 \ov r^3 \sin^2 \theta  } \de_\varphi \phi^3_3 - {2 \ov r^3} \Big( 
\de_\theta  + 3 \cot \theta  \Big) \phi^3_2;
\end{eqnarray}
\begin{eqnarray}
\De^2 \phi^3_2 & = & \lt[ {r - 2m \ov r} \de^2_r + {1 \ov r^2} 
\de^2_\theta  + {1 \ov r^2 \sin^2 \theta  } \de^2_\varphi + {2r - 3m \ov 
r^2} \de_r + {3 \cot \theta \ov r^2} \de_\theta \ri. \nonumber\\& & \mbox{} 
\lt. - {3r - 4m \ov r^3} - {3 \cot^2 \theta  \ov r^2} \ri] \phi^3_2 + 
{2(r - 2m) \ov r^2 \sin^2 \theta  } \de_\varphi \phi^2_1 \nonumber\\& & 
\mbox{} + {2(r - 2m)\ov r^2} 
\de_\theta  \phi^3_1 + {2\cot \theta  \ov r^2 \sin^2 \theta  } 
\de_\varphi \phi^2_2 - {2\cot \theta  \ov r^2 \sin^2 \theta  } 
\de_\varphi \phi^3_3.
\end{eqnarray}
While for a vector:
\begin{eqnarray}
\De^2 v_1 & = & \lt[ {r - 2m \ov r} \de^2_r -{{\hat{L}}^2 \ov r^2} 
+ {2r - m \ov r^2} \de_r - {2(r - 2m) \ov r^3} \ri] v_1 \nonumber\\& & 
\mbox{} - {2 \ov r^3 } \Big[ \de_\theta  + \cot \theta  \Big] 
v_2 - {2 \ov r^3 \sin^2 \theta  } \de_\varphi v_3;
\end{eqnarray}
\begin{eqnarray}
\De^2 v_2 & = & \lt[ {r - 2m \ov r} \de^2_r -{{\hat{L}}^2 \ov r^2} 
+ {m \ov r^2} \de_r - {r - m \ov r^3} - {1 \ov r^2} \cot^2 \theta  \ri] 
v_2 \nonumber\\& & \mbox{} + {2 (r - 2m) \ov r^2 } \de_\theta  
v_1 - {2 \cot \theta  \ov r^2 \sin^2 \theta  } \de_\varphi v_3;
\end{eqnarray}
\begin{eqnarray}
\De^2 v_3 & = & \lt[ {r - 2m \ov r} \de^2_r -{{\hat{L}}^2 \ov r^2} 
+ {m \ov r^2} \de_r + {m \ov r^3} - {2 \ov r^2} \cot \theta  \de_\theta  
\ri] v_3 \nonumber\\& & \mbox{} + {2 (r - 2m) \ov r^2 } \de_\varphi 
v_1 + {2 \cot \theta  \ov r^2} \de_\varphi v_2. 
\label{vettre}
\end{eqnarray}
Here with $\hat{L}$ we mean the usual operator of angular momentum 
expressed in polar coordinate. According to the decomposition
of Regge-Wheeler given in the Appendix B, we obtain:

\noindent{\it The even part}

Substituting the even part in the Appendix B into 
eqs.\reff{dedphi} -...\reff{vettre}; taking into account the rules of 
commutation, we find:
\begin{eqnarray}
& & \lt[ - {\hat{L}}^2, {1 \ov \sin \theta } \Big( \de_\theta - \cot 
\theta \Big) \ri] Y_{lm} (\theta, 
\varphi ) = 2 {\cot \theta \ov \sin \theta } \Big[ - {\hat{L}}^2 - 2 
\de^2_\theta \Big] Y_{lm} (\theta, \varphi ) \nonumber\\& & \mbox{} 
~~~~ + {4 \ov \sin^3 \theta } \Big[ \Big( \de_\theta - \cot \theta 
\Big) Y_{lm} (\theta , \varphi ) \Big],
\label{cuuno}
\end{eqnarray}
\beq
\lt[ - {\hat{L}}^2, {1 \ov \sin \theta } \ri] Y_{lm} (\theta, 
\varphi ) = - 2 {\cot \theta \ov \sin \theta } \de^2_\theta Y_{lm} (\theta 
, \varphi ) + {1 \ov \sin^3 \theta } Y_{lm} (\theta , \varphi ) ,
\label{cddue}
\eeq
\beq
\Big[ - {\hat{L}}^2, \de_\theta \Big] Y_{lm} (\theta, 
\varphi ) = (1 + \cot^2 \theta ) \de_\theta Y_{lm} (\theta 
, \varphi ) + 2 \cos \theta \de^2_\varphi {Y_{lm} (\theta , \varphi ) 
\ov \sin^3 \theta },
\label{cttre}
\eeq
\beq
\Big[ - {\hat{L}}^2, \sin \theta \de_\theta \Big] Y_{lm} (\theta, 
\varphi ) = - 2 \cot \theta \sin \theta \lp Y_{lm} (\theta 
, \varphi ).
\label{cqqua}
\eeq
Defining,
\beq
D_l \equiv N^2 \de^2_r + {2r - 3m \ov r^2} \de_r - {\lp \ov r^2}
\eeq
we easily obtain:
\begin{eqnarray}
\De^2 \Phi^1_1 & = & \lt\{ \lt[ D_l - 4 {r - 2m \ov r^3} \ri] H(r) + 4 
{r - 2m \ov r^2} \lp K(r) \ri. \nonumber\\& & \mbox{} \lt. + 2 {r - 2m \ov 
r^3} \Big[ 2G_1 (r) - G_2 (r) \lp \Big] \ri\} Y_{lm} (\theta , \varphi 
),
\end{eqnarray}
\begin{eqnarray}
\De^2 \Phi^2_2 & = & \lt\{ \lt[ D_l - 2 {r - 2m \ov r^3} \ri] \Big[ 
G_1 (r) + G_2 (r) \de^2_\theta \Big] \ri. \nonumber\\& & \mbox{} ~~ + 2 
{r - 2m \ov r^3} H(r) + 4 
{r - 2m \ov r^2} K(r) \de^2_\theta \nonumber\\& & \mbox{} \lt. + 
2 {G_2 (r) \ov r^2} \Big[ \lp + 2 \de^2_\theta \Big] \ri\} Y_{lm} 
(\theta , \varphi ),
\end{eqnarray}
\begin{eqnarray}
\De^2 \Phi^3_3 & = & \lt\{ \lt[ D_l - 2 {r - 2m \ov r^3} \ri] \Big[ 
G_1 (r) - G_2 (r) \Big( \lp + \de^2_\theta \Big) \Big] \ri. 
\nonumber\\& & \mbox + 4 {r - 2m \ov r^2} K(r) \lt[ {1 \ov \sin^2 
\theta } \de^2_\varphi + \cot \theta \de_\theta \ri] + 2 
{r - 2m \ov r^3} H(r) \nonumber\\& & \mbox{} \lt. - {2 \ov r^2} G_2 
(r) \Big[ \lp + 2 \de^2_\theta \Big] \ri\} Y_{lm} (\theta , \varphi ),
\end{eqnarray}
\begin{eqnarray}
\De^2 \Phi^2_1 & = & \lt\{ \lt[ D_l + 2 {r - m \ov r^2} \de_r - {2r - 
9m \ov r^3} \ri] K(r) + {2 \ov r^3} H(r) - {2 \ov r^3} G_1(r) \ri. 
\nonumber\\& & \mbox{} \lt. + {2 \ov r^3} \Big[ \lp + 1 \Big] G_2 (r) 
\ri\} \de_\theta Y_{lm} (\theta , \varphi ),
\end{eqnarray}
\begin{eqnarray}
\De^2 \Phi^3_1 & = & \lt\{ \lt[ D_l + 2 {r - m \ov r^2} \de_r - {2r - 
9m \ov r^3} \ri] K(r) + {2 \ov r^3} H(r) - {2 \ov r^3} G_1(r) \ri. 
\nonumber\\& & \mbox{} \lt. + {2 \ov r^3} \Big[ \lp + 1 \Big] G_2 (r) 
\ri\} {1 \ov \sin^2 \theta } \de_\varphi Y_{lm} (\theta , \varphi ),
\end{eqnarray}
\begin{eqnarray}
\De^2 \Phi^3_2 & = & \lt\{ \lt[ D_l + 2 {r + 2m \ov r^3} \ri] G_2 (r) 
+ 4 {r - 2m \ov r^2} K(r) \ri\} {1 \ov \sin^2 \theta } \Big( 
\de_\theta \nonumber\\& & \mbox{} - \cot \theta \Big) \de_\varphi 
Y_{lm} (\theta , \varphi ),
\end{eqnarray}
\begin{eqnarray}
\De^2 v_1 & = & \lt\{ \lt[ {r - 2m \ov r} \de^2_r - {\lp \ov r^2} + {2 
r - m \ov r^2} \de_r - 2 {r - 2m \ov r^3} \ri] U(r) \ri. 
\nonumber\\& & \mbox{} \lt. + {2 \ov r^3} \lp V(r) \ri\} \de_\theta 
Y_{lm} (\theta , \varphi ),
\end{eqnarray}
\begin{eqnarray}
\De^2 v_2 & = & \lt\{ \lt[ {r - 2m \ov r} \de^2_r - {\lp \ov r^2} + {m 
\ov r^2} \de_r + {m \ov r^3} \ri] C(r) \ri. 
\nonumber\\& & \mbox{} \lt. + 2{r - 2m \ov r^2} B(r) \ri\} \de_\theta 
Y_{lm} (\theta , \varphi ),
\end{eqnarray}
\begin{eqnarray}
\De^2 v_3 & = & \lt\{ \lt[ {r - 2m \ov r} \de^2_r - {\lp \ov r^2} + {m 
\ov r^2} \de_r + {m \ov r^3} \ri] C(r) \ri. 
\nonumber\\& & \mbox{} \lt. + 2{r - 2m \ov r^2} B(r) \ri\} \de_\varphi 
Y_{lm} (\theta , \varphi ).
\end{eqnarray}

\noindent{\it The Odd part}

Using eqs.\reff{cuuno}, \reff{cddue}, \reff{cttre}, \reff{cqqua} 
and the expressions of the odd part in the Appendix B, we obtain:
\beq
\De^2 \Phi^1_1 = 0,
\eeq
\begin{eqnarray}
\De^2 \Phi^2_2 & = & \lt[ \lt( D_l + 2 {r + 2m \ov r^3} \ri) F_2 (r) - 
4 {r - 2m \ov r^2} F_1 (r) \ri] \nonumber\\& & \mbox{} {1 \ov \sin 
\theta } \Big[ \de_\theta - \cot \theta \Big] \de_\varphi Y_{lm} 
(\theta , \varphi ),
\end{eqnarray}
\begin{eqnarray}
\De^2 \Phi^3_3 & = & - \lt[ \lt( D_l + 2 {r + 2m \ov r^3} \ri) F_2 (r) - 
4 {r - 2m \ov r^2} F_1 (r) \ri] \nonumber\\& & \mbox{} {1 \ov \sin 
\theta } \Big[ \de_\theta - \cot \theta \Big] \de_\varphi Y_{lm} 
(\theta , \varphi ),
\end{eqnarray}
\begin{eqnarray}
\De^2 \Phi^2_1 & = & \lt[ \lt( D_l + 2 {r - m \ov r^2} \de_r - {2r - 
9m \ov r^3} \ri) F_1 (r) - {1 \ov r^3} \Big[ \lp - 2 \Big] F_2 (r) \ri] 
\nonumber\\& & \mbox{} \cdot \lt[ - {1 \ov \sin \theta } \de_\varphi Y_{lm} 
(\theta , \varphi ) \ri],
\end{eqnarray}
\begin{eqnarray}
\De^2 \Phi^3_1 & = & \lt[ \lt( D_l + 2 {r - m \ov r^2} \de_r - {2r - 
9m \ov r^3} \ri) F_1 (r) - {1 \ov r^3} \Big[ \lp - 2 \Big] F_2 (r) \ri] 
\nonumber\\& & \mbox{} \cdot {1 \ov \sin \theta } \de_\varphi Y_{lm} 
(\theta , \varphi ), 
\end{eqnarray}
\begin{eqnarray}
\De^2 \Phi^3_2 & = & \lt[ \lt( D_l + 2 {r + 2m \ov r^3} \ri) F_2 (r) - 
4 {r - 2m \ov r^2} F_1 (r) \ri] {1 \ov 2 \sin \theta } 
\nonumber\\& & \mbox{} \cdot \lt[ {1 \ov \sin^2 \theta } \de^2_\varphi 
+ \cot \theta  \de_\theta - \de^2_\theta \ri] Y_{lm} (\theta , \varphi ),
\end{eqnarray}
\beq
\De^2 v_1 = 0, 
\eeq
\beq
\De^2 v_2 \lt[ {r - 2m \ov r} \de^2_r - {\lp \ov r^2} + {m \ov r^2} 
\de_r + {m \ov r^3} \ri] D(r) \lt( - {1 \ov \sin \theta } \de_\varphi 
Y_{lm} (\theta , \varphi ) \ri),
\eeq
\beq
\De^2 v_3 \lt[ {r - 2m \ov r} \de^2_r - {\lp \ov r^2} + {m \ov r^2} 
\de_r + {m \ov r^3} \ri] D(r) \lt( {1 \ov \sin \theta } \de_\varphi 
Y_{lm} (\theta , \varphi ) \ri).
\eeq

\section{Explicit form of eigenequations}

The eigenequations are given by:
\begin{eqnarray}
& & N^2 \lt[ - \De^2 \Phi^b_a + 3 \delta^s_1 \Phi^b_{a|s} {m \ov r^2} 
- \delta^s_1 \Phi^{~~|b}_{as} {m \ov r^2} - \delta^s_1 \Phi^b_{s|a} {m 
\ov r^2} + \Phi^b_c R^c_a \ri. \nonumber\\& & \mbox{} ~~~~ \lt. + 
\Phi^c_a R^b_c - {8 m^2 \ov r^3 (r - 2m)} \delta^1_{(a} h^{b)}_1 \ri] 
+ v^b_{~|a} + v^{~|b}_a + \eta^b_a \tau = 4 \lambda \Phi^b_a
\end{eqnarray}
where $\tau$ is the scalar function explicitly expressed in eq.\reff{gi}.
Analogous to the decomposition of Regge-Wheller, we separate 
the radial part from the angular part.

\subsection{Equations for the even part:}

\begin{enumerate}
\item[$\bullet$] Equations of eigen-vectors.
\begin{enumerate}
\item[ ] Substituting the symbols of Christoffel and 
tensor of Ricci indicated in Appendix A into the explicit formulas 
of $\Phi^b_a$ and $v_a$ in Appendix B for the even part, and using 
the formula calculated in C.1, we calculate:
\item[-] the equation for $H$ ($a=b=1$):
\begin{eqnarray}
& & N^2 \lt\{ \lt[ {r - 2m \ov r} \de^2_r + 2 {r - 2m \ov r^2} \de_r - 
{2 \ov 3} {6 r^2 - 27 mr + 23 m^2 \ov r^3(r - 2m)} \ri] H(r) \ri. 
\nonumber\\& & \mbox{} ~~ + 4 {r - 
2m \ov r^2} \lp K(r) \nonumber\\& & \mbox{} ~~ \lt. + 2 {r - 2m 
\ov r^3} \Big[ 2 G_1 (r) - \lp G_2 (r) \Big] \ri\} Y_{lm} (\theta , 
\varphi ) \nonumber\\& & \mbox{} = \lt\{ - H(r) 4 \lambda + {4 \ov 3} 
\lt[ {r - 2m \ov r} \de_r U(r) - {r - 3m \ov r^2} U(r) + {V(r) \ov 2} 
{\lp \ov r^2} \ri] \ri. \nonumber\\& & \mbox{} ~~ \lt. + N^2 {\lp \ov 
r^2} H(r) \ri\} Y_{lm} (\theta , \varphi ).
\end{eqnarray}
\item[-] the equation for the combine function $2G_1 
(r) - \lp G_2 (r)$ ($a = b = 2$ an then $a = b = 3$):
\begin{eqnarray}
& & N^2 \lt\{ \lt[ {r - 2m \ov r} \de^2_r + 2 {r - 2m \ov r^2} \de_r - 
{2 \ov r^2} \ri] \Big( 2 G_1 (r) \ri. \nonumber\\& & \mbox{} ~~ - \lp G_2 (r) \Big) + {4 \ov 3} {3r^2 - 12mr +7m^2 \ov r^3(r - 2m)} H(r) 
\nonumber\\& & \mbox{} ~~ \lt. 
- 4 {2r - 3m \ov r^2} \lp K(r) \ri\} Y_{lm} (\theta , 
\varphi ) \nonumber\\& & \mbox{} = \lt\{ - 4 \lambda \Big[ 2G_1 (r) - 
\lp G_2 (r) \Big] \ri. \nonumber\\& & \mbox{} ~~ + N^2 {\lp \ov r^2} \Big[ 
2G_1 (r) - \lp G_2 (r) \Big] \nonumber\\& & \mbox{} ~~ - {4 \ov 3} \lt[ 
{r - 2m \ov r} \de_r U(r) - {r - 3m \ov r^2} U(r) \ri. \nonumber\\& & \mbox{} 
~~ \lt. \lt. + {1 \ov 2} {V(r) \ov r^2} \lp \ri] 
\ri\} Y_{lm} (\theta , \varphi ).
\end{eqnarray}
\item[-] the equation for $G_2 (r)$($a = b = 2$ an then $a = b = 2$):
\begin{eqnarray}
& & N^2 \lt\{ \lt[ {r - 2m \ov r} \de^2_r + {4r - 7m \ov r^2} \de_r + 
{2 \ov r^2} \ri] G_2 (r) + 2 {2r - 3m \ov r^2} K(r) \ri\} \nonumber\\& 
& \mbox{} ~~~~ \cdot \Big[ 2 \de^2_\theta + \lp \Big] Y_{lm} (\theta , 
\varphi ) \nonumber\\& & \mbox{} ~~ = \lt\{ - 4 \lambda 2G_2 (r) 
+ 2 {V(r) \ov r^2} + N^2 {\lp \ov r^2} G_2 (r) \ri\} \nonumber\\& & \mbox{} 
~~~~ \cdot \Big[ 2 \de^2_\theta + \lp \Big] Y_{lm} (\theta , \varphi ).
\end{eqnarray}
\item[-] the equation for $K(r)$ ($b = 3$ e $a = 1$):
\begin{eqnarray}
& & N^2 \lt\{ \lt[ {r - 2m \ov r} \de^2_r + {4r - 7m \ov r^2} \de_r + 
{2 \ov r^2} \ri] G_2 (r) + 2 {2r - 3m \ov r^2} K(r) \ri\} \nonumber\\& 
& \mbox{} ~~~~ \cdot \Big[ 2 \de^2_\theta + \lp \Big] {1 \ov \sin^2 
\theta} \de_\varphi Y_{lm} (\theta , 
\varphi ) \nonumber\\& & \mbox{} ~~ = \lt\{ - 4 \lambda 2G_2 (r) 
+ 2 {V(r) \ov r^2} + N^2 {\lp \ov r^2} G_2 (r) \ri\} \nonumber\\& & \mbox{} 
~~~~ \cdot \Big[ 2 \de^2_\theta + 
\lp \Big] {1 \ov \sin^2 \theta } \de_\varphi Y_{lm} (\theta , \varphi ).
\end{eqnarray}
\item[-] Finally for $b = 3$ and $ a = 2$ we obtain the equation per $G_2$:
\begin{eqnarray}
& & N^2 \lt\{ \lt[ {r - 2m \ov r} \de^2_r + 2 {r - 2m \ov r^2} \de_r - 
{2 \ov r^2} \ri] \Big( 2 G_1 (r) - \lp G_2 (r) \Big) \ri. \nonumber\\& 
& \mbox{} ~~~~ \lt. + {4 \ov 3} {3r^2 - 12mr +7m^2 \ov r^3(r - 2m)} H(r) 
- 4 {2r - 3m \ov r^2} \lp K(r) \ri\} \nonumber\\& & \mbox{} ~~~~ \cdot 
{1 \ov \sin^2 \theta } \Big( 
\de_\theta - \cot \theta \Big) \de_\varphi Y_{lm} (\theta , 
\varphi ) \nonumber\\& & \mbox{} ~~ = \lt\{ - 4 \lambda \Big[ 2G_1 (r) - 
\lp G_2 (r) \Big] + N^2 {\lp \ov r^2} \Big[ 2G_1 (r) \ri. \nonumber\\& & 
\mbox{} ~~~~ - \lp G_2 (r) \Big] - {4 \ov 3} \lt[ {r - 2m \ov r} 
\de_r U(r) - {r - 3m \ov r^2} U(r) \ri. \nonumber\\& & \mbox{} ~~~~ 
\lt. \lt. + {1 \ov 2} {V(r) \ov r^2} \lp \ri] 
\ri\} {1 \ov \sin^2 \theta } \Big( 
\de_\theta - \cot \theta \Big) \de_\varphi Y_{lm} (\theta , \varphi ).
\end{eqnarray}
\end{enumerate}
From this equation, we get the eq.(\ref{10}) for the radial part.

\item[$\bullet$] Equations of constrain are,
\[ \Phi^i_i = 0 \phantom{cs scc} {\hbox {and}} \phantom{pippi} \lt( 
{\Phi^i_j \ov N} \ri)_{|i} = 0. \]
The second equation corresponds to 
\beq
\Phi^i_{j|i} = {m \ov r (r - 2m)} \Phi^1_j. 
\label{phiiji}
\eeq
The traceless condition leads to:
\[ \Big[ 2 G_1 (r) - G_2 (r) \lp + H(r) \Big] Y_{lm} (\theta , \varphi 
) \]
which coincides with \reff{gau}.

For imposing the condition of transversality, we use commutating rules 
shown in paragraph C.1.
\begin{enumerate}
\item[-] Imposing the condition \reff{phiiji} for $j = 1$, one gets:
\beq
\lt[ \de_r H(l) + {3r - 7m \ov r( r - 2m)} H(r) - K(r) \lp \ri] Y_{lm} 
(\theta , \varphi ) = 0
\eeq
\item[-] if $j = 2$ one gets:
\begin{eqnarray}
& & \Big\{ r(r - 2m) \de_r K(r) + 4 (r - 2m) K(r) + G_1 (r) 
\nonumber\\& & \mbox{} ~~~~ + G_2 (r) \Big[ 1 - \lp \Big] \Big\} 
\de_\theta Y_{lm} (\theta , \varphi ) = 0
\end{eqnarray}
\item[-] Finally, if $j = 3$, one gets:
\begin{eqnarray}
& & \Big\{ r(r - 2m) \de_r K(r) + 4 (r - 2m) K(r) + G_1 (r) 
\nonumber\\& & \mbox{} ~~~~ + G_2 (r) \Big[ 1 - \lp \Big] \Big\} 
\de_\varphi Y_{lm} (\theta , \varphi ) = 0,
\end{eqnarray}
\end{enumerate}
\end{enumerate}
and one verifies that the boundary conditions \reff{pa} and 
\reff{cre} are equivalent to:
\begin{eqnarray}
& & \{ r^2 [ H(r) U(r) + \lp K(r) V(r) ] \} \mid_{r = + \infty} 
\nonumber\\& & \mbox{} ~~~~ = \{ 
r^2 [ H(r) U(r) + \lp K(r) V(r) ] \} \mid_{r = 2m} = 0,
\end{eqnarray}
\beq
\left\{ \begin{array}{l}
U(r) \rightarrow O\lt( {1 \ov r} \ri) \nonumber\\ 
\phantom{U(r) \rightarrow O\lt( {1 \ov r} \ri) }~~ \phantom{pippi 
pippi} r \rightarrow + \infty \nonumber\\
V(r) \rightarrow O(1).
\end{array} \right.
\eeq

\subsection{Equations for the odd part}

\begin{enumerate}
\item[$\bullet$] Eigenequation:\\
\begin{enumerate}
\item[-] Putting $a = b = 1$, one gets identity $1 \equiv 1$.
\item[-] Putting $a = b = 2$, one gets the equation for $F_2 (r)$:
\begin{eqnarray}
& & N^2 \lt[ \lt( D_l - 3{m \ov r^2} \de_r + {2 \ov r^2} \ri) F_2 (r) 
\ri. \nonumber\\& & \mbox{} ~~ \lt. 
+ 2 {2r - 3m \ov r^2} F_1 (r) \ri] {1 \ov \sin \theta } \nonumber\\& & 
\mbox{} ~~~~ \cdot \Big[ \de_\theta - \cot \theta \Big] \de_\varphi 
Y_{lm} (\theta , \varphi ) \nonumber\\& & \mbox{} = \lt[ - 4 \lambda 
F_2 (r) + {1 \ov r^2} D(r) \ri] {1 \ov \sin \theta } \Big[ \de_\theta 
\nonumber\\& & \mbox{} ~~~~ - \cot \theta \Big] \de_\varphi Y_{lm} 
(\theta , \varphi ).
\label{stars}
\end{eqnarray}
\item[-] Putting $a = b = 3$ one gets equation \reff{stars}. 
\item[-] Putting $a = 1$ and $b = 2$ one gets:
\begin{eqnarray}
& & N^2 \lt[ \lt( D_l + 2{r - 2m \ov r^2} \de_r - {2 \ov r^2} {r^2 - 
5mr + 5 m^2 \ov r - 2m} \ri) F_1 (r) \ri. \nonumber\\& & \mbox{} ~~~~ 
~~ 
\lt. + {1 \ov r^3} \Big(2 - \lp \Big) F_2 (r) \ri] \Big[ - {1 \ov \sin 
\theta} \de_\varphi Y_{lm} (\theta , \varphi ) \Big] \nonumber\\& & 
\mbox{} ~~ = \lt[ - 4 \lambda 
F_1 (r) - {1 \ov r^2} \lt( \de_r - {2 \ov r} \ri) D(r) \ri] 
\nonumber\\& & \mbox{} ~~~~ ~~ \cdot \Big[ - {1 \ov 
\sin \theta} \de_\varphi Y_{lm} (\theta , \varphi ) \Big] .
\end{eqnarray}
\item[-] Putting $a = 1$ and $b = 2$ one gets:
\begin{eqnarray}
& & N^2 \lt[ \lt( D_l + 2{r - 2m \ov r^2} \de_r - {2 \ov r^2} {r^2 - 
5mr + 5 m^2 \ov r - 2m} \ri) F_1 (r) \ri. \nonumber\\& & \mbox{} ~~~~ 
~~
\lt. + {1 \ov r^3} \Big(2 - \lp \Big) F_2 (r) \ri] \Big[ {1 \ov \sin 
\theta} \de_\theta Y_{lm} (\theta , \varphi ) \Big] \nonumber\\& & 
\mbox{} ~~ = \lt[ - 4 \lambda 
F_1 (r) - {1 \ov r^2} \lt( \de_r - {2 \ov r} \ri) D(r) \ri] 
\nonumber\\& & \mbox{} ~~~~ ~~ \Big[ {1 \ov 
\sin \theta} \de_\theta Y_{lm} (\theta , \varphi ) \Big] .
\end{eqnarray}
\item[-] Putting $a = 2$ and $b = 3$ one gets:
\begin{eqnarray}
& & N^2 \lt[ \lt( D_l - 3{m \ov r^2} \de_r + {2 \ov r^2} \ri) F_2 (r) 
+ 2 {2r - 3m \ov r^2} F_1 (r) \ri] {1 \ov 2\sin \theta } \nonumber\\& & 
\mbox{} ~~~~ \cdot \lt[ {1 \ov \sin^2 \theta} \de^2_\varphi + \cot \theta 
\de_\theta - \de^2_\theta \ri] Y_{lm} (\theta , \varphi ) \nonumber\\& 
& \mbox{} ~~ = \lt[ - 4 \lambda 
F_2 (r) + {1 \ov r^2} D(r) \ri] {1 \ov 2\sin \theta } \lt[ {1 \ov 
\sin^2 \theta} \de^2_\varphi \ri. \nonumber\\& & \mbox{} \lt. 
\phantom{{1 \ov sin}} + \cot \theta 
\de_\theta - \de^2_\theta \ri] Y_{lm} (\theta , \varphi ). 
\end{eqnarray}
\end{enumerate}
These give \reff{raval}, \reff{aval}, which are used for the conditions 
of gauge. The traceless condition is automatically satisfied, while 
the \reff{phiiji} gives:
\begin{enumerate}
\item[-] Per $j = 1$ identity $0 \equiv 0$.
\item[-] Per $j = 2$:
\begin{eqnarray}
& & \lt\{ - \de_r F_1 (r) r(r - 2m) - 4(r - 2m) F_1 (r) \phantom{{1 
\ov r}} \ri. \nonumber\\& & \mbox{} ~~~ \lt. + F_2 (r) \lt[ 
1 - {\lp \ov 2} \ri] \ri\} {1 \ov \sin 
\theta } \de_\varphi Y_{lm} (\theta , \varphi ) = 0.
\end{eqnarray}
\item[-] Per $j = 3$:
\begin{eqnarray}
& & \lt\{ \de_r F_1 (r) r(r - 2m) + 4(r - 2m) F_1 (r) \phantom{{1 \ov 
2}} \ri. \nonumber\\& & \mbox{} ~~~~ \lt. - F_2 (r) \lt[ 
1 - {\lp \ov 2} \ri] \ri\} {1 \ov \sin 
\theta } \de_\varphi Y_{lm} (\theta , \varphi ) = 0.
\end{eqnarray}
\end{enumerate}
Finally the boundary conditions \reff{pa} e \reff{cre} become:
\begin{eqnarray}
& & \Big[ \lp r^2 F_1 (r) D(r) \Big]_{r = + \infty} \nonumber\\& & 
\mbox{} ~~~~ = \Big[ \lp r^2 F_1 (r) D(r) \Big]_{r = 2m} = 0
\end{eqnarray}
and
\beq
D_1 (r) \rightarrow O(1) ~~~~ {\hbox {per}} ~~~~ r \rightarrow + 
\infty .
\eeq
\end{enumerate}

\section{Appendix D: Normalization of eigenfunctions}

According to the scalar product \reff{scp}, one needs to normalize
the quantity:
\begin{eqnarray}
Q & = & \int_\Sigma dr d\theta d\varphi {r^2 \sin \theta \ov \lt( 1 - 
{2m \ov r} \ri) } \Big[ {\Phi^1_1}^* (r, \theta ,\varphi ) \Phi^1_1 
(r, \theta ,\varphi ) + {\Phi^2_2}^* (r, \theta ,\varphi ) \Phi^2_2 
(r, \theta ,\varphi ) \nonumber\\& & \mbox{} {\Phi^3_3}^* (r, \theta , 
\varphi ) \Phi^3_3 (r, \theta ,\varphi ) + 2r (r - 2m) {\Phi^2_1}^* (r, 
\theta ,\varphi ) \Phi^2_1 (r, \theta ,\varphi ) \nonumber\\& & 
\mbox{} + 2r(r - 2m) \sin^2 \theta {\Phi^3_1}^* (r, \theta ,\varphi ) 
\Phi^3_1 (r, \theta ,\varphi ) \nonumber\\& & \mbox{} + 2 \sin^2 \theta 
{\Phi^3_2}^* (r, \theta ,\varphi ) \Phi^3_2 (r, \theta ,\varphi ) \Big] .
\end{eqnarray}
We introduce:
\[ K = K(l); ~~ H = H(r); ~~ G_1 = G_1 (r); ~~ G_2 = G_2 (r); ~~ F_1 = 
F_1 (r) \] 
and
\[ F_2 = F_2 (r); Y = Y_{lm} (\theta , \varphi ) . \]
We indicate, 
\[ Y_{lm} (\theta , \varphi ) = {e^{im \varphi} \ov \sqrt{2 \pi}} {1 
\ov C_{lm}} P^m_l (\cos \theta ) (-1)^{{|m| - m \ov 2}} \]
where
\[ C_{lm} = \sqrt{{2 \ov 2l + 1} {(l + m)! \ov (l - m)!}} , \]
\[ P^m_l (\cos \theta ) = \sin^{|m|} \theta {d^{|m|} \ov d(\cos \theta 
)^{|m|}} P_l (\cos \theta ) \]
\[ P_l (\cos \theta ) = {1 \ov 2^l l!} {d^l \ov d \cos^l \theta } 
\sin^{2l} \theta , \]
with the conversion
\[ \int^\pi_0 d\theta \sin \theta  P_{lm} (\cos \theta ) P_{lm} (\cos 
\theta ) = C^2_{lm} \]
and 
\[
   \de_\varphi Y = i m Y, \phantom{pippi pippi} \de_\varphi Y^* = - 
   i m Y^* . 
\]
The even part is given:
\begin{eqnarray}
Q^{(+)} & = & \int_\Sigma dr d\theta d\varphi {r^2 \sin \theta \ov 1 - {2m 
\ov r}} \Big[ {\Phi^1_1}^* \Phi^1_1 + 2r(r - 2m) \Big( {\Phi^2_1}^* 
\Phi^2_1 + \sin^2 \theta {\Phi^3_1}^* \Phi^3_1 \Big) \nonumber\\& & 
\mbox{} ~~ + 2\sin^2 \theta {\Phi^3_2}^* \Phi^3_2 + {\Phi^3_3}^* \Phi^3_3 
\Big] \nonumber\\& & \mbox{} = \int_Z dr d\theta d\varphi {r^2 \sin 
\theta \ov 1 - {2m \ov r}} \lt[ H^2 Y^* Y + 2r(r - 2m) \lt( K^2 
\de_\theta Y^* \de_\theta Y \phantom{{a \ov e}} \ri. \ri. \nonumber\\& & 
\mbox{} ~~ \lt. + {K^2 \ov \sin^2 \theta } \de_\varphi Y^* 
\de_\varphi Y \ri) \nonumber\\& & \mbox{} ~~ + 2 G^2_2 {1 \ov 
\sin^2 \theta } \Big[ (\de_\theta - \cot \theta ) \de_\varphi Y^* 
\Big] \Big[ (\de_\theta - \cot \theta ) \de_\varphi Y \Big] 
\nonumber\\& & \mbox{} + (G_1 + 
G_2 \de^2_\theta ) Y^* (G_1 + G_2 \de^2_\theta )Y \nonumber\\& & 
\mbox{} ~~ + \lt( G_1 + G_2 {1 \ov \sin^2 \theta } \de^2_\varphi 
+ G_2 \cot \theta \de_\theta \ri) Y^* \lt( G_1 \phantom{{e \ov b}} 
\ri. \nonumber\\& & \mbox{} ~~ \lt. \lt. + G_2 {1 \ov \sin^2 
\theta } \de^2_\varphi + G_2 \cot \theta \de_\theta \ri) Y \ri] 
\nonumber\\& & \mbox{} = \int^\infty_{2m} dr {r^2 \ov 1 - {2m \ov r}} 
H^2 + \int^\infty_{2m} dr {r^2 \ov 1 - {2m \ov r}} K^2 \lp 2r(r - 2m) 
\nonumber\\& & \mbox{} ~~ + \int_\Sigma dr d\theta d\varphi {r^2 \sin 
\theta \ov 1 - {2m \ov r}} \lt[ 2 G^2_2 \de_\theta \lt( {1 \ov \sin 
\theta } \de_\varphi Y^* \ri) \ri. \nonumber\\& & \mbox{} ~~~~ \lt. 
\cdot \de_\theta \lt( {1 \ov \sin \theta } 
\de_\varphi Y \ri) + 2 G^2_1 Y^* Y \ri. \nonumber\\& & \mbox{} ~~ + 
\hat{L} Y^* \hat{L} Y G^2_2 + 2 G_1 Y^* G_2 \hat{L} Y - 2 G^2_2 
\de^2_\theta Y^* \nonumber\\& & \mbox{} ~~~~ \lt. \cdot \lt( {1 \ov 
\sin^2 \theta } \de^2_\varphi + \cot \theta \de_\theta \ri) Y \ri] .
\end{eqnarray}
As a result, we have:
\begin{eqnarray}
Q^{(+)} & = & \int^\infty_{2m} dr {r^2 \ov N^2} \lt\{ H^2 + \lp K^2 
2r(r - 2m) \phantom{{1 \ov 2}} \ri. \nonumber\\& & \mbox{} + 2G^2_1 + 
l^2 (l + 1)^2 {G_2}^2 - 2 \lp G_1 G_2 \nonumber\\& & \mbox{} + 
2{G_2}^2 \int d\theta d\varphi 
\sin \theta \lt[ \de_\theta \lt( {1 \ov \sin \theta } \de_\varphi Y^* 
\ri) \de_\theta \lt( {1 \ov \sin \theta } \de_\varphi Y \ri) \ri.
\nonumber\\& & \lt. \lt. \mbox{} - 
\de^2_\theta Y^* \lt( {1 \ov \sin \theta } \de^2_\varphi + \cot \theta 
\de_\theta \ri) Y \ri] \ri\} .
\label{eqp}
\end{eqnarray}
We calculate:
\begin{eqnarray}
& & \int d\varphi d \theta \sin \theta \mid \de_\theta \Big( {1 \ov 
\sin \theta } \de_\varphi Y \Big) \mid^2 \nonumber\\& & \mbox{} ~~ = 
{m^2 \ov {C_{lm}}^2} \int^1_{-1} dx \lt\{ (1 - x^2) \lt[ \de_x \lt( (1 
- x^2)^{{m - 1 \ov 2}} \de^m P_l (x) \ri) \ri]^2 \ri\} \nonumber\\& & 
\mbox{} ~~ = {m^2 \ov {C_{lm}}^2} \int^1_{-1} dx (1 - x^2) \Big[ - (m - 
1) x (1 - x^2)^{{m - 1 \ov 2} - 1} \de^m P_l (x) \nonumber\\& & \mbox{} 
~~ ~~~~ + (1 - x^2)^{{m - 1 \ov 2}} \de^{m + 1} P_l (x) \Big]^2 
\nonumber\\& & \mbox{} ~~ = {m^2 \ov 
{C_{lm}}^2} \int^1_{-1} dx \Big[ (m - 1)^2 x^2 (1 - x^2)^{m - 2} \de^m 
P_l (x) \de^m P_l (x) \nonumber\\& & \mbox{} ~~ ~~~~ + (1 - x^2)^m \de^{m + 
1} P_l (x) \de^{m + 1} 
P_l (x) \nonumber\\& & \mbox{} ~~~~ ~~  - (m - 1) x ( 1 - x^2)^{m - 1} 
\de^m P_l (x) \de^{m + 1} P_l (x) \Big] \nonumber\\& & \mbox{} ~~ 
= \Big[ - (m - 1)^2 J_{(m, l)} + (m - 1)^2 K_{(m, l)} \nonumber\\& & \mbox{} 
~~~~ ~~ + J_{(m + 1, l)} - 
2(m -1) H_{(m , l)} \Big] {m^2 \ov {C_{lm}}^2}
\label{effuno}
\end{eqnarray}
where
\beq
\left\{ \begin{array}{l}
J_{(m, l)} \equiv \int^1_{-1} (1 - x^2)^{m - 1} \de^m P_l (x) \de^m 
P_l (x) dx \nonumber\\
K_{(m, l)} \equiv \int^1_{-1} (1 - x^2)^{m - 2} \de^m P_l (x) \de^m 
P_l (x) dx \nonumber\\
H_{(m , l)} \equiv \int^1_{-1} x(1 - x^2)^{m - 1} \de^m P_l (x) \de^{m 
+ 1} P_l (x) dx \nonumber\\
I_{(m , l)} \equiv - \int^1_{-1} x(1 - x^2)^m \de^m P_l (x) \de^{m 
+ 1} P_l (x) dx .
\end{array} \right.
\label{effea}
\eeq
And:
\begin{eqnarray}
& & \int d \theta d \varphi \sin \theta \lt[ - \de^2_\theta Y^* \lt( 
{1 \ov \sin^2 \theta } \de^2_\varphi + \cot \theta \de_\theta \ri) Y 
\ri] \nonumber\\& & \mbox{} ~~ = \int d \theta d \varphi \sin \theta \lt[ 
\lp Y^* \lt( - {1 \ov \sin^2 \theta } m^2 Y + \cot \theta \de_\theta Y 
\ri) \ri. \nonumber\\& & \mbox{} ~~~~ ~~ \lt. + m^4 { 1 \ov \sin^4 \theta} 
Y^* Y + \cot^2 \theta \de_\theta Y^* \de\theta Y - 2m^2 {\cot 
\theta \ov \sin^2 \theta } Y^* \de_\theta Y \ri] \nonumber\\& & 
\mbox{} ~~ = \Big\{ \lp \Big[ - m^2 J_{(m, l)} + I_{(m, l)} - m 
{C_{lm}}^2 + m J_{(m, l)} \Big] \nonumber\\& & \mbox{} ~~~~ ~~ + \Big[ 
m^4 K_{(m, l)} + J_{(m +1, l)} - m^2 J_{(m, l)} - 2m H_{(m, l)} + m^2 
K_{(m, l)} \nonumber\\& & \mbox{} ~~~~ ~~ - {C_{l, m+1}}^2 + m^2 
{C_{lm}}^2 - m^2 J_{(m, l)} - 2m I_{(m, l)} 
\nonumber\\& & \mbox{} ~~~~ ~~ + 2m^2 H_{(m, l)} + 2m^3 J_{(m, l)} - 
2m^3 K_{(m, l)} \Big] \Big\} {1 \ov 2}.
\label{effedue}
\end{eqnarray}
Summing \reff{effuno} and \reff{effedue}, ones gets\footnote{We omit 
the index $l$.}:
\begin{eqnarray}
A_{(m, l)} & = & {\lp \ov {C_{lm}}^2} \Big[ J_m m (1 - m) + I_m - 
{C_{lm}}^2 \Big] \nonumber\\& & \mbox{} + {1 \ov {C_{lm}}^2} \Big[ 2m^2 (m 
- 1)^2 K_m - {C_{l, m + 1}}^2 + m^2 {C_{lm}}^2 \nonumber\\& & \mbox{} 
+ J_m m^2 (m - 1)(3 - m) + (1 + m^2) J_{m + 1} \nonumber\\& & \mbox{} 
- 2m(m - 1)^2 H_m - 2 m I_m \Big] .
\end{eqnarray}
We calculate explicitly the integral \reff{effea} and obtain:
\begin{eqnarray}
I_m & = & - I_m + {C_{lm}}^2 - 2m \int^1_{-1} (1 - x^2)^{m - 1} x^2 
\de^m P_l (x) \de^m P_l (x) dx \nonumber\\& & \mbox{} = -I_m + (2m + 
1) {C_{lm}}^2 - 2m J_m
\end{eqnarray}
thus,
\beq
J_m = {1 \ov m} I_m +{\lt( m + {1 \ov 2} \ri) \ov m} {C_{lm}}^2 .
\eeq
And 
\[ (1 - x^2) \de^{m + 1} P_l (x) = 2m x \de^m P_l (x) - (l + m)(l -m + 
1) \de^{m - 1} P_l (x) \]
therefore
\begin{eqnarray}
& I_m & = - {1 \ov 2m} {C_{lm}}^2 - {(l + m)(l - m + 1) \ov 2m} 
\int^1_{-1} (1 - x^2)^m \de^{m - 1} P_l (x) \de^{m + 1} P_l (x) dx 
\nonumber\\& & \mbox{} = - {1 \ov 2m} \Big[ {C_{l, m+ 1}}^2 - (l + m) 
(l - m + 1) {C_{l, m}}^2 \nonumber\\& & \mbox{} - (l + m) (l - m + 1) 
2m I_{m - 1} \Big] ;
\end{eqnarray}
owing to ${C_{l, m+1}}^2 = {C_{l, m}}^2 (l + m + 1) (l - m)$ e ${C_{l, 
m}}^2 = {C_{l, m-1}}^2 (l + m) (l - m + 1)$, one has:
\begin{eqnarray}
I_m & = & {C_{l, m}}^2 + (l + m) (l - m + 1) I_{m - 1} \nonumber\\& & 
\mbox{} = 2 {C_{l, m}}^2 + (l + m) (l + m - 1) (l - m + 1) (l - m + 2) 
I_{m - 2} \nonumber\\& & \mbox{} = m^2 {C_{l, m}}^2 + {(l + m)! \ov (l 
- m)!} I_0.
\end{eqnarray}
\begin{eqnarray}
I_0 & = & - \int^1_{-1} x P_l (x) \de P_l (x) dx = - I_0 + \int^1_{-1} 
[P_l]^2 dx - x P^2_l (x) \mid^1_{-1} \nonumber\\& & \mbox{} = {1 \ov 
2l+1} - 1.
\end{eqnarray}
As a result $I_m = (m - l) {C_{l, m}}^2$ e $J_m = {2l + 1 \ov 2m} {C_{l, 
m}}^2$. Now we have:
\begin{eqnarray}
& J_m & = - \int^1_{-1} \de^{m - 1} P_l (x) \de \Big[ (1 - x^2)^{m -1} 
\de^m P_l \Big] dx \nonumber\\& & \mbox{} = 2(m - 1)  \int^1_{-1} 
\de^{m - 1} P_l (x) \de^m P_L (x) x (1 - x^2)^{m - 2} dx \nonumber\\& 
& \mbox{} ~~~~ - \int^1_{-1} \de^{m -1} P_l (x) \de^{m + 1} P_l (x) (1 
- x^2)^{m - 1} dx \nonumber\\& & \mbox{} = 2(m - 1) H_{m - 1} - 
\int^1_{-1} P_l (x) \de^{m +1} P_l (x) (1 - x^2)^{m -1} dx \nonumber\\& & 
\mbox{} = 2(m - 1) H_{m - 1} - \Big[ \int^1_{-1} \de^{m -1} P_l (x) 
2mx \de^m P_l (x) (1 - x^2)^{m - 2} dx \nonumber\\& & \mbox{} ~~~~ - 
(l + m)(l - m + 1) \int^1_{-1} \de^{m - 1} P_l (x) \de^{m - 1} P_l (x) 
(1 - x^2)^{m - 2} dx \Big]
\nonumber\\& & \mbox{} = - 2H_{m - 1} + (l + m) (l - m + 1) J_{m - 1}
\end{eqnarray}
and then
\beq
H_m = {(l + m + 1)(l - m) \ov 2} J_m - {J_{m + 1} \ov 2}
\eeq
or
\beq
H_m = {(2l + 1) (l + m +1) (l - m) \ov 4m (m + 1)}.
\eeq
Finally
\begin{eqnarray}
J_m & = & \int^1_{-1} (1 - x^2)^{m + 1} \de^m P_l (x) \de^m P_l (x) dx 
\nonumber\\& & \mbox{} = -2 \int^1_{-1} x (1 - x^2)^{m - 1} \de^{m + 
1} P_l (x) \de^m P_l (x) dx \nonumber\\& & \mbox{} ~~~~ + 2(m - 1) 
\int^1_{-1} x^2 (1 - x^2)^{m - 2} \de^m P_l (x) \de^m P_l (x) 
\nonumber\\& & \mbox{} = -2H_m - 2 (m - 1) J_m + 2(m -1) K_m
\end{eqnarray}
which leads to:
\begin{eqnarray}
K_m & = & {2m - 1 \ov 2(m - 1)} J_m + {H_m \ov m - 1} \nonumber\\& & 
\mbox{} = {2l + 1 \ov 4m(m - 1)} \lt[ 2m - 1 + {(l + m + 1)(l - m) \ov 
m + 1} \ri] .
\end{eqnarray}
Therefore we find:
\begin{eqnarray}
& A_{(l, m)} & = \lp \lt[ - {2l + 1 \ov 2} (m - 1) + m - l - m \ri] + 
{1 \ov 2} m(m - 1) (2m \nonumber\\& & \mbox{} ~~ - 1) (2l + 1) + {1 
\ov 2} {m(m - 1) (2l + 1) (l + m + 1) (l - m) \ov m + 1} \nonumber\\& 
& \mbox{} ~~ - (l + m + 1) 
(l - m) + m^2 + {1 \ov 2} m(m - 1) (3 - m) (2l + 1) \nonumber\\& & 
\mbox{} ~~ + {1 \ov 2(m + 1)} (2l + 1) (l + m + 1) (l - m) (m^2 + 1) 
\nonumber\\& & \mbox{} ~~ - 
{1 \ov 2(m + 1)} (m - 1)^2 (l + m + 1) (l - m) (2l + 1) \nonumber\\& & 
\mbox{} = - {1 \ov 2} \lp .
\end{eqnarray}
Thus one gets:
\begin{eqnarray}
& Q^{(+)} & = \displaystyle\int^{+ \infty}_{2m} \displaystyle{r^2 \ov N^2} 
\Big\{ H^2 + \lp K^2 2r (r - 2m) + 2{G_1}^2 \nonumber\\& & \mbox{} - 2 
\lp G_1 G_2 + {G_2}^2 \Big[ l^2 (l + 1)^2 - \lp \Big] \Big\} dr
\end{eqnarray}
Concerning the odd part, one gets:
\begin{eqnarray}
Q^{(-)} & = & \int_\Sigma dr d\theta d\varphi \sin \theta {r^2 \ov 
N^2} \lt\{ 2r(r - 2m) \lt[ {F_1}^2 {1 \ov \sin^2 \theta } \de_\varphi Y^* 
\de_\varphi Y \ri. \ri. \nonumber\\& & \mbox{} ~~ \lt. + {F_1}^2 
\de_\theta Y^* \de_\theta Y + 2 {F_2}^2 {1 \ov \sin^2 \theta } \mid 
(\de_\theta - \cot \theta ) \de_\varphi Y_{lm} \mid^2 \ri] \nonumber\\& & 
\mbox{} ~~ \lt. + {1 \ov 2} {F_2}^2 \mid \Big[ \lp + {2 \ov \sin^2 \theta } 
\de^2_\varphi + 2 \cot \theta \de_\theta \Big] Y \mid^2 \ri\} 
\nonumber\\& & \mbox{} = \int^\infty_{2m} dr {r^2 \ov N^2} \lt\{ 
2r(r - 2m) {F_1}^2 (r) \lp + {F_2}^2 \lt[ {l^2 (l + 1)^2 \ov 2} \ri. 
\ri. \nonumber\\& & \mbox{} ~~ + 2 \lt( m^2 \int d\omega \lt( 
\mid \de_\theta \Big( {1 \ov \sin \theta} Y \Big) \mid^2 + \lp \Big( - 
{m^2 \ov \sin^2 \theta } Y^* Y \ri. \ri. \nonumber\\& & \mbox{} ~~ + 
\cot \theta \de_\theta Y^* Y \Big) \Big( {m^4 \ov 
\sin^4 \theta } Y^* Y + \cot^2 \theta  \de_\theta Y^* \de_\theta Y 
\nonumber\\& & \mbox{} ~~ \lt. \lt. \lt. \lt. - 
{m^2 \ov \sin^2 \theta} \cot^2 \theta \de_\theta Y^* \de_\theta Y 
\Big) \ri) \ri) \ri] \ri\} .
\end{eqnarray}
Using the results from calculating the even part, we obtain:
\beq
Q^{(-)} = \int^\infty_{2m} {r^2 \ov N^2} \lt\{ 2r(r - 2m) \lp {F_1}^2 
(r) + {F_2}^2 \lt[ {l^2 (l + 1)^2 \ov 2} - \lp \ri] \ri\} dr
\label{eqm}
\eeq

\section{Appendix E: Study of eigenequations in the limit of flat space $m=0$}

At the first we study the even part. The eigenequations and
the equations of constrains becomes:
\beq
\left\{ \begin{array}{l}
\lt[ \de^2_r + \displaystyle{2 \ov r} \de_r - \displaystyle{\lp \ov 
r^2} - \displaystyle{4 \ov r^2} \ri] H(r) + \displaystyle{4 \ov r} \lp 
K(r) \nonumber\\ \phantom{pippi} + \displaystyle{2 \ov r^2} \Big( 2G_1 
(r) - \lp G_2 (r) \Big) = - 4\displaystyle{\lambda \ov \hbar^2} H(r) 
\nonumber\\
\\
\lt[ \de^2_r + \displaystyle{2 \ov r} \de_r - \displaystyle{\lp \ov 
r^2} + \displaystyle{2 \ov r^2} \ri] G_2(r) + \displaystyle{4 \ov r} 
K(r) = - 4\displaystyle{\lambda \ov \hbar^2} G_2(r) \nonumber\\
\\
\lt[ \de^2_r + \displaystyle{2 \ov r} \de_r - \displaystyle{\lp \ov 
r^2} - \displaystyle{2 \ov r^2} \ri] G_1 (r) + \displaystyle{2 \ov r^2} 
G_2 (r) \lp \nonumber\\ \phantom{pippi} + \displaystyle{2 \ov r^2} H(r) 
= - 4\displaystyle{\lambda \ov \hbar^2} G_1 (r) \nonumber\\
\\
\lt[ \de^2_r + \displaystyle{2 \ov r} \de_r - \displaystyle{\lp \ov 
r^2} - \displaystyle{2 \ov r^2} \ri] K(r) + \displaystyle{2 \ov r^3} 
H(r) - \displaystyle{2 \ov r^3} \Big[ G_1 (r) \nonumber\\ \phantom{pippi} 
+ \Big( 1 - \lp \Big) G_2 (r) \Big) \Big]= - 4\displaystyle{\lambda \ov 
\hbar^2} K(r) \nonumber\\
\end{array} \right.
\label{fpu}
\eeq
and
\beq
\left\{ \begin{array}{l}
2G_1 (r) - G_2 (r) \lp H(r) = 0 \nonumber\\
\\
\de_r H(r) + 3 \displaystyle{H(r) \ov r} - K(r) \lp = 0 \nonumber\\
\\
r^2 \de_r K(r) + 4r K(r) + G_2 (r) \Big[ 1 - \lp \Big] = 0.
\end{array} \right.
\label{fpd}
\eeq
We show that eigenequations are consistent with equations of 
constrains. Substituting the first three 
equations of system \reff{fpu} into the first equation of \reff{fpd}:
\begin{eqnarray}
& & - {4\lambda \ov \hbar^2} \Big[ 2G_1 (r) - \lp G_2 (r) + H(r) \Big] 
= \lt[ \de^2_r + {2 \ov r} \de_r - {\lp \ov r^2} \ri] \nonumber\\& & 
\mbox{} ~~~~ ~~ \cdot \Big[ 2G_1 (r) - \lp G_2 (r) + H(r) \Big] .
\label{deltauno}
\end{eqnarray}
Now we substitute the first and last equations of \reff{fpu} into the
second equation of \reff{fpd}; take into account the relations:
\[
   \de_r \lt[ {2 \ov r} \de_r H(r) \ri] = {2 \ov r} \de_r \Big[ \de_r 
   H(r) \Big] - {2 \ov r^2} \de_r H(r)
\]
\[
  \de_r \lt[ {2 \ov r^2} \de_r H(r) \ri] = {1 \ov r^2} \de_r \Big[ \de_r 
   H(r) \Big] - {2 \ov r^3} \de_r H(r)
\]
\[ {1 \ov r} \Big[ \de^2_r H(r) \Big] = \de^2_r \lt[ {1 \ov r} H(r) 
   \ri] + {2 \ov r^2} \de_r H(r) - {2 \ov r^3} H(r)
\]
\[
   {1 \ov r} \lt[ {1 \ov r} \de_r H(r) \ri] = {1 \ov r} \de_r \lt[ {1 
   \ov r} H(r) \ri] + {1 \ov r^3} H(r)
\]
and using the third equation of \reff{fpd}, we obtain:
\begin{eqnarray}
& & {4\lambda \ov \hbar^2} \lt[ \de_r H(r) + {3H(r) \ov r} - \lp K(r) 
\ri] \nonumber\\& & \mbox{} ~~~~ ~~ = \lt[ \de^2_r + {2 \ov r} \de_r - 
\lp - {2 \ov r^2} \ri] \nonumber\\& & \mbox{} ~~~~ ~~ \cdot \lt[ \de_r 
H(r) + {3 \ov r} H(r) - \lp K(r) \ri].
\label{deltaunob}
\end{eqnarray}
Finally, substituting the last three equations of system \reff{fpu} 
into the last equation of \reff{fpd}, we have
\[
   r^2 \de_r \Big[ \de^2_r K(r) \Big] = \de^2_r \Big[ r^2 \de_r K(r) 
   \Big] - 2 \de_r K(r) - 4 r \de^2_r K(r)
\]
\[
   r^2 \de_r \lt[ {1 \ov r} \de_r K(r) \ri] = {1 \ov r} \Big[ r^2 
   \de_r K(r) \Big] - 3 \de_r K(r)
\]
\[
   r^2 \de_r \lt[ {1 \ov r^2} K(r) \ri] = {1 \ov r^2}  \Big[ r^2 \de_r 
   K(r) \Big] - {2 \ov r} K(r)
\]
\begin{eqnarray*}
r \de^2_r K(r) & = & \de^2_r \Big[ r K(r) \Big] - 2 \de_r K(r) \\& & 
\mbox{} = {1 \ov r} \de_r \Big[ r^2 \de_r K(r) \Big] - 2 \de_r K(r)
\end{eqnarray*}
\[
   r \lt[ {1 \ov r} \de_r K(r) \ri] = {1 \ov r} \de_r \Big[ r K(r) 
   \Big] - {1 \ov r} K(r)
\]
and the same for \reff{fpd}, one gets:
\begin{eqnarray}
& & - {4 \lambda \ov \hbar^2} \Big[ r^2 \de_r K(r) + 4 r K(r) + G_1 
(r) + G_2 (r) \Big( 1 - \lp \Big) \Big] \nonumber\\& & \mbox{} ~~~~ ~~ 
= \lt[ \de^2_r - {\lp \ov r^2} \ri] \Big[ r^2 \de_r K(r) + 4rK(r) 
\nonumber\\& & \mbox{} ~~~~ ~~ + G_1 (r) + G_2 (r) \Big( 1 - \lp \Big) 
\Big].
\label{deltatrec}
\end{eqnarray}
The equation \reff{deltauno}, \reff{deltaunob} and \reff{deltatrec} 
show that the conditions of constrain \reff{fpd} are compatible with 
the eigenequation \reff{fpu}. Thus, we can solve the eigenequation
by using the first equation of \reff{fpu} together with the eq.\reff{fpd}. 
As a result, we have an equation for $H(r)$ only:
\beq
\lt[ \de^2_r + {6 \ov r} \de_r + {6 - \lp \ov r^2} \ri] H(r) = 4 
{\lambda \ov \hbar^2} H(r).
\label{pft}
\eeq
Introducing the variable $\rho = 2 {\sqrt{\lambda} \over \hbar} r$ and 
indicating `` ' '' as the derivative with respect to new variable, one 
gets:
\beq
H'' + {6 \ov \rho} H' + \lt[ 1 - {(l + 3) (l - 2) \ov \rho^2} \ri] = 
0.
\label{fpq}
\eeq
Setting\footnote{ $J_\nu$ are the Bessel functions.}
\[
   H = \rho^\alpha J_\nu (\rho ),
\]
one gets:
\beq
H' = \alpha \rho^{\alpha - 1} J_\nu (\rho ) + \rho^\alpha {J'}_\nu (\rho ) 
\eeq
\beq
H'' = \alpha (\alpha - 1) \rho^{\alpha - 2} J_\nu (\rho ) + 2 \alpha 
\rho^{\alpha - 1} {J'}_\nu (\rho ) + \rho^\alpha {J''}_\nu (\rho ), 
\eeq
and substituting it into \reff{fpq}:
\begin{eqnarray}
& & \rho^\alpha \lt[ {J''}_\nu (\rho ) + {2 \alpha + 6 \ov \rho} {J'}_\nu 
+ \lt( 1 \phantom{{1 \ov h}} \ri. \ri. \nonumber\\& & \mbox{} ~~~~ ~~ \lt. 
\lt. - {(l + 3)(l - 2) - 6 \alpha - \alpha (\alpha - 1) \ov \rho^2} 
\ri) J_\nu (\rho ) \ri] = 0 
\end{eqnarray}
with $2 \alpha + 6 = 1$, so that the above equation is the equation of 
Bessel function with $\nu^2 = \lt( l + {1 \ov 2} \ri)^2$.
Finally,
\beq
H(r) = C r^{-{5 \ov 2}} J_{l + {1 \ov 2}} \lt(2 {\sqrt{\lambda} \ov 
\hbar} r \ri) .
\eeq
From the second equation of \reff{fpd}, one finds:
\begin{eqnarray}
K(r) & = & {c \ov lp} \lt[ {r^{- 7/2} \ov 2} J_{l + {1 \ov 2}} \lt( 2 
{\sqrt{\lambda} \ov \hbar} r \ri) \ri. \nonumber\\& & \mbox{} \lt. - 
{2 \sqrt{\lambda} \ov \hbar} {1 \ov r^{5 / 2}} {J'}_{l + {1 \ov 2}} {2 
\sqrt{\lambda} r \ov \hbar} \ri] .
\end{eqnarray}
Recalling the relation for the Bessel function
\[
   {d \ov dz} \Big( z^{- \nu} J_\nu (z) \Big) = - z^{- \nu} J_{\nu + 
   1} (z),
\]
we have
\beq 
{J'}_{l + {1 \ov 2}} \lt( 2 {\sqrt{\lambda} \ov \hbar} r \ri) = {\hbar 
\ov 2 \sqrt{\lambda}} r \lt( l + {1 \ov 2}\ri) J_{l + {1 \ov 2}} \lt( 2 
{\sqrt{\lambda} \ov \hbar} r \ri) - J_{l + {3 \ov 2}} \lt( 2 
{\sqrt{\lambda} \ov \hbar} r \ri), 
\eeq
thus
\beq
K(r) = {c \ov \lp} \lt[ {l + 1 \ov r^{7/2}} J_{l + {1 \ov 2}} \lt( 2 
{\sqrt{\lambda} r \ov \hbar} \ri) - {2 \sqrt{\lambda} \ov \hbar} {1 \ov 
r^{5 / 2}} J_{l + {3 \ov 2}} \lt( 2 {\sqrt{\lambda} r \ov \hbar} \ri) 
\ri] .
\eeq
From the first equation of \reff{fpd} one finds:
\beq
2G_1 (r) - G_2 (r) \lp + c r^{-5/2} J_{l + {1 \ov 2}} \lt( 2 
{\sqrt{\lambda} r \ov \hbar} \ri) = 0.
\eeq
From the last, instead, one find:
\begin{eqnarray}
& & G_1 (r) + G_2 (r) \Big[ 1 - \lp \Big] = - {4c \ov \lp } \lt\{ {l+ 
1 \ov r^{5/2}} J_{l + {1 \ov 2}} \lt( 2 {\sqrt{\lambda} \ov \hbar} 
\ri) \ri. \nonumber\\& & \mbox{} ~~~~ ~~ - 2 {\sqrt{\lambda} \ov \hbar 
r^{3/2} J_{l + {3 \ov 2}}} \lt( 2 {\sqrt{\lambda} r \ov \hbar} \ri) - 
{c \ov \lp} \lt[ - {7 \ov 2} {l + 1 \ov r^{5/2}} J_{l + {1 \ov 2}} 
\lt(2 {\sqrt{\lambda} r \ov \hbar} \ri) \ri. \nonumber\\& & \mbox{} 
~~~~ ~~ + {l + 1 \ov r^{3/2}} 2 {\sqrt{\lambda} \ov \hbar} {J'}_{l + 
{1 \ov 2}} \lt(2 {\sqrt{\lambda} r \ov \hbar} \ri) + {5 \ov 2} {2 
\sqrt{\lambda} \ov \hbar r^{3/2}} \lt(2 {\sqrt{\lambda} r \ov \hbar} \ri) 
\nonumber\\& & \mbox{} ~~~~ ~~ \lt. \lt. - {4 \ov r^{1/2}} {\lambda 
\ov \hbar^2} {J'}_{l + {3 \ov 2}} \Big( 2 \sqrt{\lambda} r \Big) \ri] 
\ri\} \nonumber\\& & \mbox{} ~~ = - {c \ov l \lt( l + {1 \ov 2} \ri) } 
\lt[ { (l + 1)^2 \ov r ^{5/2}} J_{l + {1 \ov 2}} \lt(2 {\sqrt{\lambda} 
r \ov \hbar} \ri) - {2l + 3 \ov r^{3/2}} 2 {\sqrt{\lambda} \ov \hbar} 
J_{l + {3 \ov 2}} \lt(2 {\sqrt{\lambda} r \ov \hbar} \ri) \ri. 
\nonumber\\& & \mbox{} ~~~~ ~~ \lt. + {4 \ov r^{1/2}} {\lambda \ov 
\hbar^2} J_{l + {5 \ov 2}} \lt(2 {\sqrt{\lambda} r \ov \hbar} \ri) 
\ri] .
\end{eqnarray}
At this point, we can adopt approximation $\hbar\rightarrow 0$, which 
is equivalent to the limit of $J_{\nu + {1 \ov 2}} (z)$ for 
$z \rightarrow + \infty$,
\[ 
   J_{\nu + {1 \ov 2}} (z) \sim \sqrt{{2 \ov \pi}} {1 \ov \sqrt{z}} 
   \cos \lt[ z - (\nu + 1) {\pi \ov 2} \ri] ,
\]
and one has:
\beq
H(r) \sim C \sqrt{{1 \ov \pi} \hbar \sqrt{{2 \ov \lambda}}} {1 \ov 
r^3} \cos \lt[ {\sqrt{\lambda} r \ov \hbar} - \lp {\pi \ov 2} \ri]
\eeq
\begin{eqnarray}
K(r) & \sim & {C \ov \lp} \sqrt{{\hbar \ov \pi} \sqrt{{2 \ov \lambda}}} 
2 {\sqrt{\lambda} r \ov \hbar} \cos \lt[ 2 {\sqrt{\lambda} r \ov \hbar} 
- (l + 2) {\pi \ov 2} \ri] \nonumber\\& & \mbox{} = {1 \ov \lp} 2 
\sqrt{\lambda} c \sqrt{{1 \ov \pi} \sqrt{{2 \ov \lambda}}} \sin \lt[ 2 
{\sqrt{\lambda} r \ov \hbar} - \lp {\pi \ov 2} \ri]
\label{emmoche}
\end{eqnarray}
\begin{eqnarray}
2G_1 (r) - G_2 (r) \lp \sim -c \sqrt{{\hbar \ov \pi} \sqrt{{2 \ov 
\lambda}}} {1 \ov r^3} \cos \lt[ 2 {\sqrt{\lambda} r \ov \hbar} - (l + 
1) {\pi \ov 2} \ri]
\end{eqnarray}
\begin{eqnarray}
& G_1 (r) & + G_2 (r) \Big[ 1 - \lp \Big] \nonumber\\& & \mbox{} \sim 
- {c \ov \lp} \sqrt{{\hbar \ov 
\pi} \sqrt{{2 \ov \lambda}}} {4 \lambda \ov r \hbar^2} \cos \lt[ 2 
{\sqrt{\lambda} r \ov \hbar} - (l + 3) {\pi \ov 2} \ri] .
\end{eqnarray}
From the last two equations, one finds:
\[
   G_1 (r) \sim {\lp \ov 2} G_2 (r)
\]
con
\begin{eqnarray}
G_2 (r) & \sim & {1 \ov 1 - {\lp \ov 2}} {c \ov \lp} \sqrt{{\hbar \ov 
\pi} \sqrt{{2 \ov \lambda}}} \nonumber\\& & \mbox{} ~~ \cdot {4 
\lambda \ov r \hbar^2} \cos \lt[ 2 {\sqrt{\lambda} r \ov \hbar} - (l + 
1) {\pi \ov 2} \ri] .
\label{flau}
\end{eqnarray}
It is clear then that the \reff{tetd} coincide with these solutions
in the limit $m \rightarrow 0$ if $ \alpha (l) = i (l + 1) {\pi \ov 2}$.

Now we study the odd part. In this case, the equation becomes:
\beq
\left\{ \begin{array}{l}
\lt[ \de^2_r + \displaystyle{6 \ov r} \de_r - \displaystyle{\lp \ov 
r^2} + \displaystyle{6 \ov r^2} \ri] F_1 (r) = - 4 \lambda F_1 (r) 
\nonumber\\
\\
\lt[ \de^2_r + \displaystyle{2 \ov r} \de_r - \displaystyle{\lp \ov 
r^2} + \displaystyle{2 \ov r^2} \ri] F_2(r) - \displaystyle{4 \ov r} 
F_1 (r) \nonumber\\ \phantom{pippi} = - 4 \lambda F_2(r) 
\end{array} \right.
\label{fpc}
\eeq
\beq
F_2 (r) \lt( 1 - {\lp \ov 2} \ri) = \Big( r^2 \de_r + 4r \Big) F_1 
(r).
\label{fps}
\eeq
Again the conditions of constrain are compatible with the eigenequation. 
In fact one finds:
\begin{eqnarray}
& & -4 \lambda \lt[ \Big( r^2 \de_r + 4r \Big) F_1 (r) - \lt( 1 - {\lp 
\ov 2} \ri) F_2 (r) \ri] \nonumber\\& & \mbox{} ~~~~ ~~ = \Big( r^2 
\de_r + 4r \Big) \lt[ \de^2_r + {6 \ov r} \de_r - {\lp \ov r^2} + {6 
\ov r^2} \ri] F_1 (r) \nonumber\\& & \mbox{} ~~~~ ~~ - \lt[ 1 - {\lp 
\ov 2} \ri] \lt[ \de^2_r + {2 \ov r} \de_r - {\lp \ov r^2} + {2 \ov 
r^2} \ri] F_2 (r) \nonumber\\& & \mbox{} ~~~~ ~~ + \lt[ 1 - {\lp \ov 
2} \ri] {4 \ov r} F_1 (r).
\label{fdeltadue}
\end{eqnarray}
Based on:
\[
   r^2 \de_r \Big[ \de^2_r F_1 (r) \Big] = \de^2_r \Big[ r^2 \de_r F_1 
   (r) \Big] - 2 \de_r F_1 (r) - 4 r \de^2_r F_1 (r)
\]
\[
   r^2 \de_r \lt[ {1 \ov r} \de_r F_1 (r) \ri] = {1 \ov r} \de_r \Big[ 
   r^2 \de_r F_1 (r) \Big] - 3\de_r F_1 (r)
\]
\[ r^2 \de_r \lt[ {1 \ov r^2} F_1 (r) \ri] = {1 \ov r^2} \de_r \Big[ 
   r^2 \de_r F_1 (r) \Big] - {2 \ov r} F_1 (r)
\]
\[
   r \de^2_r F_1 (r) = \de^2_r \Big[ r F_1 (r) \Big] - 2 \de_r F_1 (r)
\]
\[
   r \lt[ {1 \ov r} \de_r F_1 (r) \ri] = {1 \ov r} \de_r \Big[ r F_1 
   (r) \Big] - {1 \ov r} F_1
\]
one finds 
\begin{eqnarray}
& & - 4\lambda \lt[ \Big( r^2 \de_r + 4 r \Big) F_1 (r) - \lt( 1 - 
{\lp \ov 2} \ri) F_2 (r) \ri] \nonumber\\& & \mbox{} ~~~~ ~~ = \lt[ 
\de^2_r + {2 \ov r} \de_r - {\lp \ov r^2} + {2 \ov r^2} \ri] 
\nonumber\\& & \mbox{} ~~~~ ~~ \cdot \lt\{ \Big[ r^2 \de_r + 4r \Big] 
F_1 (r) - \lt[ 1 - {\lp \ov 2} \ri] F_2 (r) \ri\}
\end{eqnarray}
which show the compatibility of constrain. From the first equation of 
\reff{fpc}, one gets:
\[
   F_1 (r) = {c \ov r^{5/2}} J_{l + {1 \ov 2}} \lt( 2 {\sqrt{\lambda} 
   r \ov \hbar} \ri) .
\]
As a consequence:
\begin{eqnarray}
& & \lt( 1 - {\lp \ov 2} \ri) F_2 (r) = c \lt[ - {5 \ov 2} {1 \ov 
r^{3/2}} J_{l + {1 \ov 2}} \lt( 2 {\sqrt{\lambda} r \ov \hbar} \ri) 
\ri. \nonumber\\& & \mbox{} ~~~~ ~~ \lt. + {2 \sqrt{\lambda} \ov 
r^{1/2} \hbar} {J'}_{l + {1 \ov 2}} \lt( 2 {\sqrt{\lambda} r \ov \hbar} 
\ri) + {4 \ov r^{3/2}} J_{l + {1 \ov 2}} \lt( 2 {\sqrt{\lambda} r \ov 
\hbar} \ri) \ri] \nonumber\\& & \mbox{} ~~~ = c \lt[ {l + 2 \ov r^{3/2}} 
J_{l + {1 \ov 2}} \lt( 2 {\sqrt{\lambda} r \ov \hbar} \ri) - { 2 
\sqrt{\lambda} r^{1/2} \ov \hbar} J_{l + {3 \ov 2}} \lt( 2 
{\sqrt{\lambda} r \ov \hbar} \ri) \ri] .
\end{eqnarray}
In the limit $\hbar \rightarrow 0$, one finds:
\beq
F_1 (r) \sim c \sqrt{{\hbar \ov \pi} \sqrt{{2 \ov \lambda}}} {1 \ov 
r^3} \cos \lt( 2 {\sqrt{\lambda} r \ov \hbar} - \lp {\pi \ov 2} \ri) 
\label{doma}
\eeq
\begin{eqnarray}
& & \lt( 1 - {\lp \ov 2} \ri) F_2 (r) \sim c \sqrt{{\hbar \ov \pi} 
\sqrt{{2 \ov \lambda}}} {1 \ov 2} {2 \sqrt{\lambda} \ov \hbar} \cos 
\lt( 2 {\sqrt{\lambda} r \ov \hbar} - (l + 2) {\pi \ov 2} \ri) 
\nonumber\\& & \mbox{} ~~~ = c \sqrt{{\hbar \ov \pi} 
\sqrt{{2 \ov \lambda}}} {1 \ov r} 2 {\sqrt{\lambda} \ov \hbar} \sin 
\lt[ 2{\sqrt{\lambda} r \ov \hbar} - \lp {\pi \ov 2} \ri) .
\label{famoni}
\end{eqnarray}
Again $\alpha (l) = -i (l + 1)$.

\end{document}